\documentclass[11pt,a4paper]{article}

\usepackage{jheppub}
\usepackage{amssymb,amsmath}

\usepackage{graphicx}
\usepackage{slashed}
\usepackage[normalem]{ulem}

\usepackage{hyperref}

 \usepackage[final]{pdfpages}

\pdfoutput=1
\usepackage{epstopdf}

\usepackage{mathrsfs}
\usepackage{psfrag}
\usepackage{color}

\usepackage{slashed}
\usepackage{simplewick}
\usepackage{cancel}
\usepackage[utf8]{inputenc}

\usepackage{amsmath,array,amsfonts,graphicx,wrapfig,arydshln,lscape,float,multirow,longtable,rotating,makecell}
\usepackage{url}

\usepackage{subcaption}

\usepackage[backgroundcolor=black!5, linecolor=green!60]{todonotes}

\hypersetup{pdftitle={},pdfcreator={},linkcolor=[rgb]{0.15,0.35,0.75},colorlinks=true,citecolor=[rgb]{0.675,0,0.2},urlcolor=[rgb]{0.15,0.35,0.65}}
\let\olditemize\itemize\renewcommand{\itemize}{\vspace{-2pt}\olditemize\setlength{\itemsep}{1pt}\setlength{\parskip}{0pt}\setlength{\parsep}{-0pt}}
\let\oldenumerate\enumerate\renewcommand{\enumerate}{\vspace{-4pt}\oldenumerate\setlength{\itemsep}{1pt}\setlength{\parskip}{0pt}\setlength{\parsep}{0pt}}

\newcommand{\p}[1]{(\ref{#1})}

\newcommand \vev [1] {\langle{#1}\rangle}
\newcommand \ket [1] {|{#1}\rangle}

\newcommand{\cN}{{\cal N}}

\newcommand{\pa}{\partial}
\newcommand{\ep}{\epsilon}

\renewcommand{\a}{\alpha}
\renewcommand{\b}{\beta}
\newcommand{\g}{\gamma}

\newcommand{\la}{\lambda}

\newcommand{\da}{{\dot\alpha}}
\newcommand{\db}{{\dot\beta}}
\newcommand{\dg}{{\dot\gamma}}

\newcommand{\tr}{\mbox{tr}}

\usepackage{soul}

\makeatletter\DeclareRobustCommand*{\bfseries}{\not@math@alphabet\bfseries\mathbf\fontseries\bfdefault\selectfont\boldmath}\makeatother

\title{\boldmath Symmetry properties of Wilson loops with a Lagrangian insertion}

\author[a]{Dmitry Chicherin} 
\author[b]{\hspace{-5pt}, Johannes M. Henn}

\affiliation[a]{LAPTh, Universit\'e Savoie Mont Blanc, CNRS, B.P. 110, F-74941 Annecy-le-Vieux, France}
\affiliation[b]{Max-Planck-Institut f\"ur Physik, Werner-Heisenberg-Institut, D-80805 M\"unchen, Germany}

\emailAdd{chicherin@lapth.cnrs.fr}
\emailAdd{henn@mpp.mpg.de}

\preprint{LAPTH-004/22, MPP-2022-19}

\abstract{
Null Wilson loops in $\mathcal{N}=4$ super Yang-Mills are dual to planar scattering amplitudes. This duality implies hidden symmetries for both objects. We consider closely related infrared finite observables, defined as the Wilson loop with a Lagrangian insertion, normalized by the Wilson loop itself. Unlike ratio and remainder functions studied in the literature, this observable is non-trivial already for four scattered particles and bears close resemblance to (finite parts of) scattering processes in non-supersymmetric Yang-Mills theory. Moreover, by integrating over the insertion point, one can recover information on the amplitude, as was recently done to compute the full four-loop cusp anomalous dimension. We study the general structure of the Wilson loop with a Lagrangian insertion, focusing in particular on its leading singularities and their (hidden) symmetry properties. Thanks to the close connection of the observable to integrands of MHV amplitudes, it is natural to expect that its leading singularities can be written as certain Grassmannian integrals. The latter are manifestly dual conformal. They also have a conformal symmetry, up to total derivatives. We find that, surprisingly, the conformal symmetry becomes an invariance in the frame where the Lagrangian insertion point is sent to infinity. Furthermore, we use integrability methods to study how higher Yangian charges act on the Grassmannian integral. We evaluate the $n$-particle observable both at tree- and at one-loop level, finding compact analytic formulas. These results are explicitly written in the form of conformal leading singularities, multiplied by transcendental functions. We then compare these formulas to known expressions for all-plus amplitudes in pure Yang-Mills theory. We find a remarkable new connection: the Wilson loop with Lagrangian insertion in $\mathcal{N}=4$ super Yang-Mills appears to predict the maximal weight terms of the planar pure Yang-Mills all-plus amplitude. We test this relationship for the two-loop $n$-point Yang-Mills amplitude, as well as for the three-loop four-point amplitude.
}

\begin{document} 
\setcounter{tocdepth}{2}
\maketitle
\flushbottom

\newpage

\section{Introduction}

Wilson lines are of fundamental importance in gauge theories. They describe many physical situations, especially in singular limits, where they provide an effective, often universal description. For example, correlation functions of Wilson loops describe the key part of infrared divergences of scattering amplitudes \cite{Almelid:2015jia}. In planar ${\cal N}=4$ super Yang-Mills (sYM), null polygonal Wilson loops even describe the full scattering amplitudes, including the finite part \cite{Alday:2007hr}. This duality implies many useful properties for both the scattering amplitudes and the Wilson loops. The former have an intrinsic (super)conformal symmetry, which is augmented by a (dual) superconformal symmetry of the Wilson loops \cite{Drummond:2007au}. Together the symmetries combine to a Yangian symmetry \cite{Drummond:2009fd}. 

Since the discovery of the Wilson loop / scattering amplitude duality countless important further developments occurred. Very remarkably, information about the planar loop integrand of the scattering amplitudes is readily available: its leading singularities are described by a Grassmannian \cite{ArkaniHamed:2009dn,Arkani-Hamed:2016byb}; it can be obtained from a loop-level version of the BCFW on-shell recursions \cite{Arkani-Hamed:2010zjl}, and there is a dual geometric description in terms of the Amplituhedron \cite{Arkani-Hamed:2013jha}. Moreover, loop integrands at high loop orders have been obtained from soft and collinear consistency relations \cite{Bourjaily:2011hi,Bourjaily:2016evz}, and via a connection to correlation functions \cite{Eden:2011we,Eden:2012tu,Ambrosio:2013pba}.

In parallel, integrated planar amplitudes were obtained to high loop orders via bootstrap methods \cite{Dixon:2011pw,Drummond:2014ffa}, relying in part on observations about the symbol alphabet \cite{Goncharov:2010jf}, novel cluster algebraic structures \cite{Golden:2013xva} and analytic properties \cite{Caron-Huot:2016owq,Caron-Huot:2019bsq} of the integrated answer (for reviews see \cite{Caron-Huot:2020bkp,Henn:2020omi}). However, apart from initial calculations used to jump-start this program, it is not known how to go directly from the integrands to the integrated answers. Filling this gap would signify important progress in our understanding and ability of performing loop calculations, and would very likely be useful beyond ${\cal N}=4$ sYM \cite{Henn:2020omi}.

So we would like to use ${\cal N}=4$ sYM as a training ground for learning how to go from the loop integrand of scattering amplitudes to the integrated answer. 
Indeed, the setting seems promising: both the planar integrand and the integrated answer are known to staggering loop orders.
However, there are two immediate concerns that should be addressed before getting started.
The first has to do with the kind of finite observables studied so far, namely ratio and remainder functions \cite{Drummond:2008vq}. 
{We use the word observable in the sense of a finite, gauge-invariant and scheme-independent quantity. For example, for scattering amplitudes, a suitably defined finite part can be identified as a particular contribution to a cross section, see e.g. \cite{Weinzierl:2011uz}.}
Unfortunately, by construction, the ratio and remainder functions are non-trivial only starting from six external particles. However, in realistic gauge theories, already the four- and five-particle scattering processes are very complicated and of enormous interest. Therefore, tools for studying such processes  directly are desirable. The second issue has to do with the fact that scattering amplitudes have infrared divergences.
The form of the latter is known in principle (and often can be given explicitly to very high loop orders), so that only the finite part of the scattering amplitudes is new. 
Moreover, one expects that simplifications occur for the finite part, while intermediate, regulated results may be considerably more complicated.
Unfortunately, despite knowing the form of the divergences, it is not in general known how to set up the calculation of the finite part without needing regularization of intermediate steps (for recent related work see \cite{Anastasiou:2018rib}).

In ref. \cite{Alday:2012hy} a new class of observables was studied that has the potential to solve these two issues.
The authors considered correlation functions of Wilson loops with local operator insertions. 
More precisely, one takes
\begin{align}\label{introdefratio}
F_n = \frac{1}{\pi^2} \frac{\langle W_n \,{\cal L}(x_0) \rangle}{ \langle W_n  \rangle}\,,
\end{align}
where $W_n$ is the $n$-sided null Wilson loop dual to scattering amplitudes, and ${\cal L}$ is the Lagrangian, inserted at an arbitrary position $x_{0}$.
We take the Wilson loop to be in the fundamental representation of the gauge group.
This quantity has a number of desirable properties.
Firstly, it is easy to see that cusp divergences of the Wilson loop (dual to infrared divergences of scattering amplitudes) drop out in the ratio in Eq. (\ref{introdefratio}).
Secondly, upon integrating over $x_0$, one recovers (the logarithmic derivative of) the logarithm of the Wilson loop itself, and via the duality, the logarithm of the planar scattering amplitude. 
Therefore, $F_n$ can be thought of as the integrand of (the logarithm of) the scattering amplitude, where all but one integrations have been carried out, in such a way as to give a finite answer. This is reminiscent of \cite{Erdogan:2011yc}, where a finite web integrand function is defined for exponentiated Wilson loops, and where divergences only come from further integrations over the web integrand. 
Thirdly, in addition to the dual conformal symmetry of the Wilson loop, one might expect $F_n$ to have further symmetries, thanks to the amplitude duality. This however is rather subtle, and is a central point of study in the present paper.

In view of the dual conformal symmetry, the ratio in (\ref{introdefratio}) is a function of $3n-11$ variables \cite{Alday:2012hy}.
In the literature, it has been mostly studied in the case $n=4$, where it is given by a non-trivial function of one variable, in analogy with four-particle scattering amplitudes.
That function is known at weak coupling to three loops, and at leading order in strong coupling \cite{Alday:2012hy,Alday:2013ip,Henn:2019swt}.

Quite remarkably, the Amplituhedron description of the planar loop integrand gives a natural way of discussing the logarithm of the amplitude (or the Wilson loop) \cite{Arkani-Hamed:2013kca}.
It can be seen that this allows one to define the integrand for $F_{n}$ directly from geometric equations, analogous to the original definition of the Amplituhedron. In this way the planar integrand for $F_n$ is arranged into a novel form, where each term is manifestly finite in four dimensions \cite{Arkani-Hamed:2021iya}. This representation is explored further for $n=4$ in \cite{Arkani-Hamed:2021iya}.

In the present paper we study the structure of the $n$-particle function $F_n$, with a particular focus on its leading singularities and their symmetry properties. Our motivation is to inform future bootstrap approaches for this quantity, and to lay the basis for future studies of anomalous symmetries.

We profit from known results for integrands of scattering amplitudes \cite{ArkaniHamed:2009dn,Arkani-Hamed:2010zjl}. More precisely, using convenient expressions for the integrand of the (logarithm of) the one- and two-loop MHV amplitudes, we are able to determine the analytic result for $F_n$ at tree-level and at one loop. We find that the leading singularities appearing in these expressions are governed by contour integrals over a Grassmannian ${\rm Gr}(2,2+n)$. The Grassmannian formula was first obtained in the context of one-loop MHV amplitudes in  \cite{ArkaniHamed:2009dn,Arkani-Hamed:2010zjl}. We find it suggestive that this formula is valid to higher loops as well. Indeed, further evidence towards this conjecture is provided by an explicit two-loop calculation of $F_5$ \cite{CHtoappear2}.

The planar loop integrand of $\mathcal{N}=4$ sYM is known to have a Yangian symmetry. However, this symmetry is somewhat formal, in the sense that it is only valid when the integrations are considered as well. For example, the leading singularities of planar amplitudes in $\mathcal{N}=4$ sYM are Yangian invariant. In our case, however, we do not integrate over the insertion point $x_0$, so that the Yangian invariance could be obscured by total derivative terms. This is not an issue for the dual conformal part of the symmetry. In our setup, it corresponds to the native symmetry of the polygonal Wilson loop, and the Lagrangian transforms covariantly under it. However, the same is not true for the conformal symmetry of amplitudes. 
 We find that the situation becomes easier if we choose a dual conformal frame where $x_0 \to \infty$, which can be done without loss of generality. In that case, the formulas for the leading singularities simplify,
 and we find that they are {\it invariant} under conformal transformations. In other words, the symmetry is manifest, as opposed to being obscured by potential total derivative terms.
We first observe this property at $n=4,5$, and then provide a proof that the Grassmannian formula is conformally invariant.

Indeed there is much encouraging work in the context of scattering amplitudes where symmetry properties could be proven starting from Grassmannian formulas, see e.g. \cite{Drummond:2010qh,Drummond:2010uq,Korchemsky:2010ut,Ferro:2015grk,Ferro:2016zmx}. 
It turns out, rather surprisingly, that the formula for the leading singularities of $F_n$ resembles very closely a special case of a Grassmannian formula for tree-level Amplituhedron volume functions considered in \cite{Ferro:2015grk,Ferro:2016zmx,Bai:2015qoa}.

 Motivated by the additional symmetry, we further explore the structure of the higher Yangian charges, using integrability methods. We do this both in the dual conformal frame $x_{0} \to \infty$, and for generic $x_0$. We find that the leading singularities are annihilated by half of the Yangian generators, and identify among them the conformal symmetry. The proof uses methods of integrability to construct Yangian generators and to study their action on the Grassmannian representation.
Specifically, we rely on constructions borrowed from the quantum inverse scattering method~\cite{Faddeev:1996iy,Sklyanin:1991ss} and integrable spin chains. Similar considerations to formulate the Yangian symmetry in QFT context has been previously applied to
the tree-level scattering amplitudes in ${\cal N}= 4$ sYM \cite{Chicherin:2013ora,Frassek:2013xza,Kanning:2014maa,Broedel:2014pia} and ABJM theory \cite{Bargheer:2014mxa}, form factors of composite operators \cite{Bork:2015fla,Frassek:2015rka}, form factors of Wilson lines \cite{Bork:2016xfn}, kernels of QCD parton evolutions \cite{Fuksa:2016tpa},  Amplituhedron volume functions \cite{Ferro:2016zmx},  splitting amplitudes \cite{Kirschner:2017vqm}, and  Fishnet diagrams \cite{Chicherin:2017cns,Chicherin:2017frs}.

Next, we analyze further the full expression for the observable $F_n$, and find that it is closely related to all-plus helicity scattering amplitudes in pure Yang-Mills theory.
In the frame $x_{0} \to \infty$, we write the leading singularities in a standard momentum-space notation,
and make contact with conformal invariants that appeared in all-plus Yang-Mills amplitudes \cite{Badger:2019djh,Henn:2019mvc}. But not only the leading singularities are the same --- in fact we find that the lowest order expression for the amplitude matches precisely that of our correlator!

Motivated by the leading order results, we compute $F_n$ for arbitrary $n$ at the one-loop order, again finding the expected structure of conformal leading singularities, multiplied by transcendental functions, which we compute explicitly.
We then
compare $F_n$ to the two-loop all-plus helicity amplitudes \cite{Dunbar:2016cxp}. 
We find evidence that the Wilson loop with the Lagrangian insertion at $L$ loops agrees with the $(L+1)$-loop all-plus amplitude in pure Yang-Mills theory! 
This is at the level of leading transcendental terms, and up to scheme-dependent terms. 
If true in general, this relationship can be used as a prediction for yet to be computed scattering amplitudes.

The outline of this paper is as follows. In section \ref{sec:recap}, we give a lightning review of key features of the Wilson loop / scattering amplitudes duality that are relevant to this paper. Section \ref{sec:defFn} is dedicated to the Wilson loop with a Lagrangian insertion. We discuss this observable at Born level, and propose Grassmannian formulas for its leading singularities at loop level. We observe that the leading singularities have a hidden, conformal symmetry, which we prove using the Grassmannian representation. In section \ref{sectionhigherorder}, we study higher order symmetries of the leading singularities. In section \ref{sec:perturbative_results} we provide evidence that we correctly identified all leading singularities of the loop corrections. In particular, we present the full analytic one-loop result for $F_n$. In section \ref{sect:all-plus-YM}, we compare $F_n$ it to the maximal-weight piece of two-loop all-plus amplitudes in pure Yang-Mills theory, and find evidence for a new duality.
We discuss future research directions in section \ref{sec:summary}.

\section{Recap of key features of the Wilson loop / scattering amplitudes duality}
\label{sec:recap}

In this section we quickly review aspects of the duality between Wilson loops and scattering amplitudes that provide useful background information for the topic developed in the paper.

Unless otherwise stated, we will be discussing the planar limit, where the `t Hooft coupling 
\begin{align}
\lambda=g_{\rm YM}^2 N_c \,,
\end{align}
stays fixed, while taking the rank of the gauge group $N_{c}$ to infinity, and letting the Yang-Mills coupling $g_{\rm YM}$ go to zero.
We will find it convenient to express results in terms of the combination
\begin{align}
    g^2 = \frac{g_{\rm YM}^2 N_c}{16 \pi^2} \,,
\end{align}
which is often used in the literature.

While there are indications that a version of the duality may hold for non-planar cases \cite{Ben-Israel:2018ckc}, it is certainly best understood in the planar limit.  For reviews, see \cite{Alday:2008yw,Henn:2020omi}. 
In its simplest form, the duality relates maximally-helicity-violating (MHV) scattering amplitudes and null polygonal Wilson loops.
On the one hand, the scattering amplitudes depend on $n$ on-shell momenta $p_{i} = \lambda_i \tilde{\lambda_i}$, obeying momentum conservation $\sum_{i=1}^{n} p_i = 0$.
On the other hand, the Wilson loops are defined on a null polygonal contour, with cusps $x_{i}$.
The duality identifies the kinematics according to
\begin{align}\label{xp}
x_{i+1} - x_{i} = p_{i+1}\,, 
\end{align}
where $x_{n+1} := x_{1}$.
The statement of the duality is that
\begin{align}\label{duality-relation}
 \langle W(x_1, \ldots ,x_n) \rangle  \, {A}_{\rm tree}^{\rm MHV}  \sim  {A}^{\rm MHV}_n  \,,
\end{align}
where $A^{\rm MHV}_n$ is the color-ordered MHV amplitude, and ${A}^{\rm MHV}_{\rm tree}$ is its tree-level value.
It is the well-known Parke-Taylor factor which ensures that both objects have the same helicity weights.
The symbol $\sim$ indicates that the objects are dual to each other, but not strictly equal. 
The reason is that the objects have different divergences:
the Wilson left-hand side of Eq.~(\ref{duality-relation}) has ultraviolet divergences (short-distance divergences arising near the cusp points) \cite{Drummond:2007au}, while the amplitude has infrared divergences (i.e. soft and collinear divergences in momentum space) \cite{Bern:2005iz}. 

It turns out that there are two ways one can sharpen the duality and replace the equivalence by an equal sign.
The first is to go to appropriately defined finite versions of the objects entering the duality. This idea has led to studying ratio and remainder functions \cite{Drummond:2008vq}.
The second is to consider both sides of eq. (\ref{duality-relation}) at the level of four-dimensional loop integrands. This version will be very useful for the purpose of this paper. 
On one side of the duality, we have the planar $ L$-loop integrand of MHV scattering amplitudes (with the tree-level factor removed). Thanks to its planarity, it can be expressed in terms of region variables, removing the ambiguity of loop momentum relabelling. On the other side of the duality, let us consider a derivative in the coupling,
\begin{align}
g^2_{\rm YM} \partial_{g^2_{\rm YM}} \langle W(x_1, \ldots ,x_n) \rangle \sim \int d^D y \, \langle W(x_1, \ldots ,x_n) {\cal L}(y)  \rangle \,.
\end{align}
Expanding this relation in the coupling, we see that integrating over a Lagrangian insertion generates one-loop corrections to the Wilson loop.
Iterating this procedure leads in a natural way to consider
\begin{align}\label{IntegrandWL}
\langle W(x_1, \ldots ,x_n)  {\cal L}(y_1) \ldots {\cal L}(y_{ L}) \rangle\,,
\end{align}
evaluated at the lowest perturbative order, as the $ L$-loop integrand of the Wilson loop. 
 
In principle, it is not obvious that the two quantities, i.e. Eq.~(\ref{IntegrandWL}) and the integrand of scattering amplitudes, should coincide, since loop integrands are defined up to total derivatives.
Indeed, taking alternative formulations of the Lagrangian can lead to such differences.
However it was noticed in refs. \cite{Eden:2012tu,Chicherin:2014uca} that when ${\cal L}$ is taken to be the chiral Lagrangian, the duality holds at integrand level.
In the following, we will define a finite observable based on the same integrands, which however involves loop integrations and therefore non-trivial transcendental functions.
It will be very useful for us to rely on insights discovered in the context of the Wilson loop / scattering amplitudes duality.

The duality implies hidden symmetries for both objects. The scattering amplitudes have a conformal symmetry, which is augmented by a dual conformal symmetry of the Wilson loop. The symmetries are anomalous at loop level but have a high predictive power: they completely fix the form of four- and five-particle scattering amplitudes, and heavily constrain higher-particle amplitudes. 
In a supersymmetrized version of the duality, both objects have a superconformal and dual superconformal symmetry \cite{Drummond:2008vq,Berkovits:2008ic}, which combine to a Yangian symmetry \cite{Drummond:2009fd}. 

Conjecturally, scattering amplitudes in ${\cal N}=4$ sYM can be written in terms of transcendental functions (of transcendental weight $2  L$ at $ L$ loops), multiplied by certain algebraic prefactors. 
The latter are computed using so-called leading singularities. Several dual formulations for computing leading singularities exist, for example in terms of residues of certain Grassmannian formulas, and in terms of plabic graphs. This helps to make the Yangian symmetry of the leading singularities manifest. Moreover, in some cases it leads to a complete classification of what form the leading singularities can take. For example, for planar MHV amplitudes, there is only a single leading singularity, the Parke-Taylor factor. This implies that the integrated answer is a linear combination of transcendental functions, with coefficients in $\mathbb{Q}$, i.e. independent of the kinematics.

\section{Null Wilson loop with a Lagrangian insertion}
\label{sec:defFn}

\subsection{Definition, kinematics, and key properties}

We define the ratio of the Wilson loop correlation function with the Lagrangian ${\cal L}$ and the vacuum expectation value of the Wilson loop $W_{F}[{\cal C}]=1/N_{F} {\rm tr}_{F} P {\rm exp}\left( i \oint_{\cal C} dx \cdot A(x) \right)$ in the fundamental representation, 
\begin{align}
F_n(x_1,\ldots,x_n;x_0) = \frac{1}{\pi^2} \frac{\vev{W_{F}[x_1,\ldots,x_n]\, {\cal L}(x_0)}}{\vev{W_{F}[x_1,\ldots,x_n]}}  \,, \label{FnBos}
\end{align}
 where the Wilson loop contour ${\cal{C}} = [x_1,\ldots,x_n]$ is an $n$-cusp polygon with the light-like edges, that is 
\begin{align}
(x_{i+1}-x_i)^2 = 0,\qquad i=1,\ldots,n,
\end{align}
embedded in Minkowski space.

There are several different forms of the $\cN = 4$ sYM Lagrangian, which all coincide modulo total derivatives and equations of motion. Here we take ${\cal L}$ to be the chiral on-shell Lagrangian. An explicit expression for the latter 
can be found in Eqs. (3.37) and (A.13) of \cite{Eden:2011yp}.
Another choice of the Lagrangian would result in the shift of $F_n$ by a total derivative term (in $x_0$).
Our choice of the Lagrangian is motivated by the fact that via the correlator/amplitude duality \cite{Eden:2010zz,Eden:2010ce,Eden:2011yp,Eden:2011ku} it exactly corresponds to the amplitude loop integrands generated by means of the loop BCFW recursion \cite{Arkani-Hamed:2010zjl}. Moreover, with this choice of the operator, ${\cal L}$ is a component of the stress-tensor supermultiplet.

A major motivation for considering the ratio \p{FnBos} is that it is a finite quantity.  
This can be understood as follows.
It is known that cusp divergences of Wilson loops exponentiate \cite{Brandt:1981kf,Korchemskaya:1992je}. 
This means that while at $ L$ loops, $\vev{W[x_1,\ldots,x_n]}$ has poles up to $\epsilon^{-2  L}$ in dimensional regularization with $D=4-2 \epsilon$, 
its logarithm has only double poles at most. 
Moreover, the exponent, $\log \vev{W}$, can be computed directly from certain web diagrams that are free of subdivergences (see \cite{Erdogan:2011yc} and references therein).
Let us now consider the following derivative,
\begin{align}\label{logderivativeW}
g^2_{\rm YM} \partial_{g^2_{\rm YM}}  \log \vev{W[x_1,\ldots,x_n]} = \int d^D x_0  \,F_n(x_1,\ldots,x_n;x_0 ) \,.
\end{align}
This means that $F_n$ can be considered as the {\it integrand} of the exponentiated Wilson loop.
We already know from the discussion of webs that $ \log \vev{W}$ is free of subdivergences. In other words the cusp divergences only arise via the final integration in $x_0$ (more precisely
from integration regions where $x_0$ approaches the light-like contour), but the integrand $F_n$ is free of divergences. 

We will therefore consider $F_n$ in $D=4$ dimensions. Note that this is not sufficient in order to use Eq.~(\ref{logderivativeW}) to obtain the full $ \log \vev{W[x_1,\ldots,x_n]}$ in dimensional regularization, 
but this is not our aim here. (Nevertheless, it was shown that this relation can be used to efficiently compute the cusp anomalous dimension, see \cite{Henn:2019swt}.) Rather we use this to motivate the finiteness of the observable $F_n$ that we study.

\begin{figure}[t]
\begin{center}
\includegraphics[width=0.3\columnwidth]{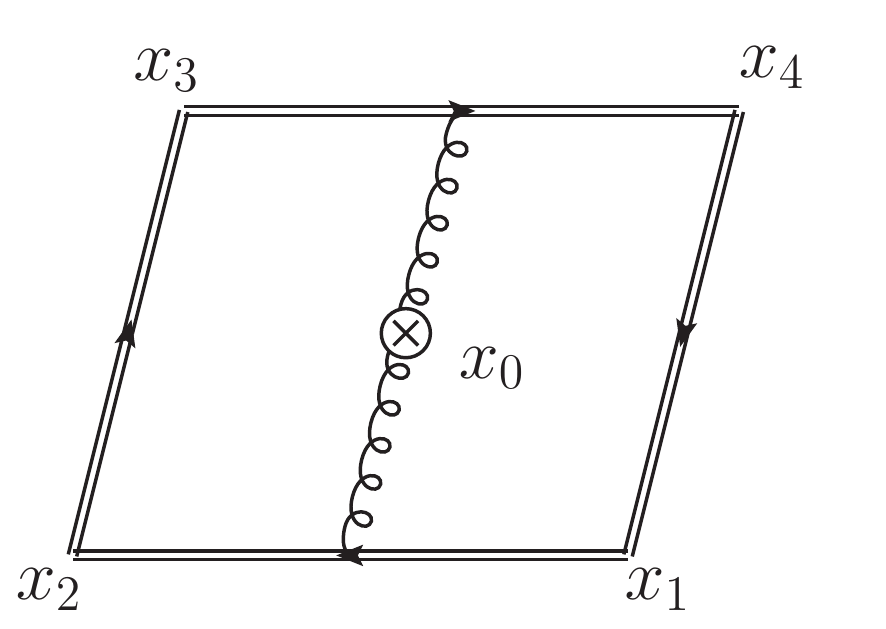}
\includegraphics[width=0.3\columnwidth]{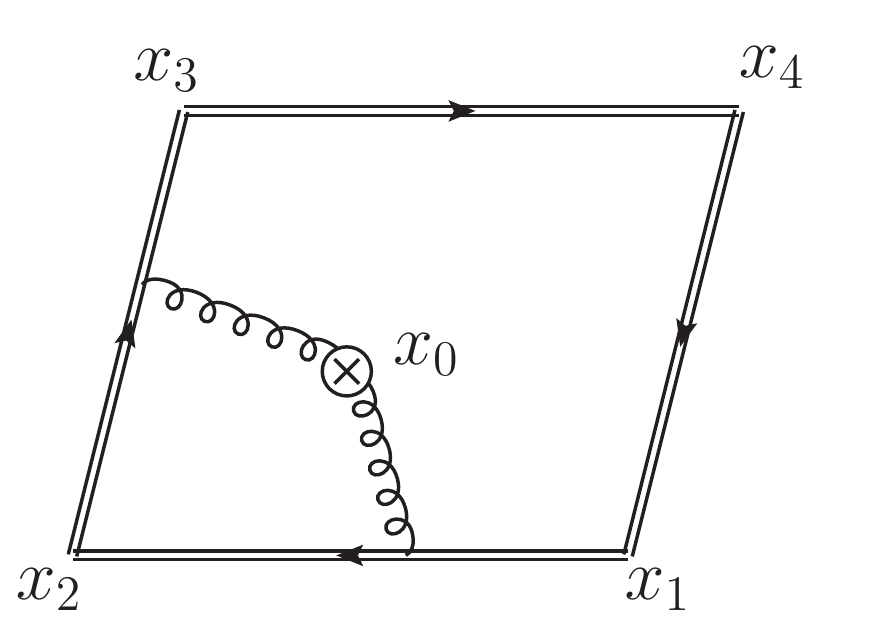}
\includegraphics[width=0.3\columnwidth]{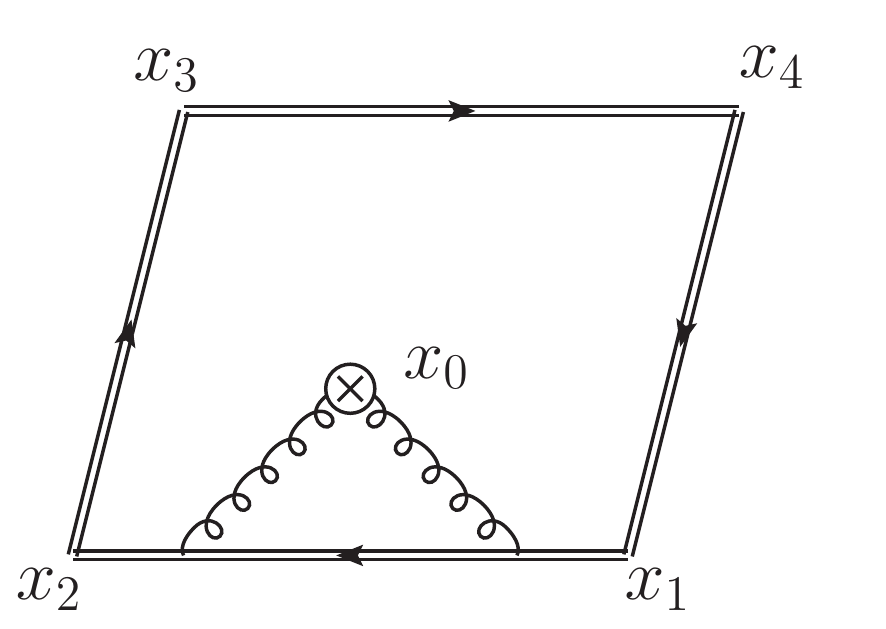}
    \caption{Representative Feynman diagrams contributing to $F_{4}^{(0)}$, i.e. at order $g_{\rm YM}^2$.}
\label{FigLinsertion4pointslowestorder}
 \end{center}
\end{figure}

Let us consider the perturbative expansion of $F_n$ in the coupling $g_{\rm YM}$.
At the lowest perturbative order, least two gluons need to be emitted from the Wilson loop contour to connect to the Lagrangian insertion (see Fig.~\ref{FigLinsertion4pointslowestorder}).
This means that the expansion starts at order ${\cal O}(g_{\rm YM}^2)$,  and we define accordingly,
\begin{align}
F_n = \sum_{ L \geq 0} (g^2)^{1+ L} F_n^{( L)} \,. \label{FnPertSer}
\end{align}
The expected uniform transcendental weight property of $\cN =4$ sYM theory suggests that the 
terms of the perturbative expansion should be of the following form
\begin{align}
F_n^{( L)} = \sum_{s} R_{n,s}\, g^{( L)}_{n,s} \label{Fnl}\,.
\end{align}
Here $\{ R_{n,s} \}$ is a set of $n$-cusp leading singularities, which are rational functions of $x_1,\ldots,x_n$ and of $x_0$,
and where the accompanying $g^{( L)}_{n,s}$ are pure functions of transcendental weight $2 L$. 
In particular, at the Born-level ($ L =0$) we have $g^{(0)}_{n,s} = 1$, and $F_n^{(0)}$ is a rational function.

For example, evaluating the Feynman diagrams of Fig~\ref{FigLinsertion4pointslowestorder}, one finds (see \cite{Eden:2010ce,Eden:2010zz} for a similar Lagrangian insertion calculation for a correlation function of scalar operators)
\begin{align}\label{Ftree4point}
F_4^{(0)} = - \frac{x_{13}^2 x_{24}^2}{x_{01}^2 x_{02}^2 x_{03}^2 x_{04}^2} \,.
\end{align}

The light-like geometry of the contour is preserved by dual-conformal transformations. The quantum corrections to the Wilson loop render the dual-conformal symmetry anomalous. In the ratio \p{FnBos}, the anomalies of the dual-conformal symmetry cancel out, and this object transforms convariantly under the dual conformal transformations. It carries zero dual-conformal weight with respect to each of the cusp points $x_1,\ldots,x_n$, and dual-conformal weight $4$ with respect to the Lagrangian point $x_0$. It is easy to see these properties for eq. (\ref{Ftree4point}), for example.

The leading singularities $ R_{n,s}$ in eq. \p{Fnl} are dual-conformal covariant, while the functions $g^{( L)}_{n,s}$ depend on dual conformal invariant cross-ratios build from $x_1,\ldots,x_n,x_0$. 
Given the kinematic setup of a null polygon, plus one additional point $x_{0}$, it turns out that there are $3 n- 11$ such dual conformal invariants \cite{Engelund:2011fg}.

Note that unlike the Wilson loop $\vev{W_n}$, $F_n$ has both a parity-even and odd part, due to the Lagrangian insertion. This means that in addition to the $3n-11$ parity-even cross ratios just mentioned, $F_n$ depends on parity-odd Lorentz invariants.
Since the square of a parity-odd variable is parity even, each parity-odd invariant can be expressed in terms of the cross-ratios, up to a sign.

For example, at five points, there is one parity-odd invariant, which we write as \cite{Ambrosio:2013pba}
\begin{align}\label{defeps012345}
 \ep_{123450} = -\frac{i}{6} \sum_{\sigma \in {\cal S}_6 } (-1)^\sigma x_{\sigma_1}^2 \, \ep(x_{\sigma_2},x_{\sigma_3},x_{\sigma_4},x_{\sigma_5})\,,
\end{align}
where $\ep(a,b,c,d)=\ep_{\mu \nu \rho \sigma} a^\mu b^\nu c^\rho d^\sigma$, $(-1)^\sigma$ stands for the signature of $\sigma$, and where the summation runs over all permutations of $\{ 1,2,3,4,5,0\}$. This expression is totally antisymmetric in the points $x_0,x_1,\ldots,x_5$. It is dual-conformally covariant with weight $-1$ with respect to each of the points \cite{Ambrosio:2013pba}. 
We remark that its square is parity even, and can be written as follows,
\begin{align}
(\ep_{123450})^2 = -\frac{1}{24} \sum_{\sigma,\tau \in {\cal S}_5} (-1)^\sigma (-1)^\tau x_{0\sigma_1}^2 x_{0\tau_1}^2 x_{\sigma_2 \tau_2}^2 
x_{\sigma_3 \tau_3}^2 x_{\sigma_4 \tau_4}^2 x_{\sigma_5 \tau_5}^2 \,, \label{epssq}
\end{align}
where the summation is over all permutations of $\{ 1,2,3,4,5\}$.

With this, we have all ingredients to write the full result for the Born-level five-particle case. According to \cite{Eden:2010zz}, it is given by 
\begin{align}
F_5^{(0)} = -
\frac{1}{2} \frac{x_{24}^2 x_{35}^2 x_{10}^2 + x_{14}^2 x_{35}^2 x_{20}^2 + x_{14}^2 x_{25}^2 x_{30}^2 + x_{13}^2 x_{25}^2 x_{40}^2 + x_{13}^2 x_{24}^2 x_{50}^2 +\ep_{123450}}{x_{10}^2 x_{20}^2 x_{30}^2 x_{40}^2 x_{50}^2} \,. \label{F50local}
\end{align}
This expression allows us to make a number of comments:
\begin{itemize}
\item The expression is manifestly dual-conformally covariant, as expected.
\item The expression is manifestly local, i.e. it has poles only at expected locations $x_{i0}^2=0$. Below we will present different representations where individual terms have spurious poles, but where a hidden conformal symmetry is manifest. 
\item The expression is a sum of a parity even and parity odd piece. One might think that the parity odd piece is unimportant, for several reasons: firstly, upon carrying out the $x_0$ integration, it vanishes, since $\log W$ is parity even; secondly, in principle it depends on the precise definition of the Lagrangian insertion. Nonetheless, we will find that it is only due to the specific combination of parity even and odd terms that $F$ has a (hidden) conformal symmetry.
\item Some readers might notice some structural similarities to all-plus amplitudes in pure Yang-Mills theory. This is indeed the case, and we will explore this connection and find hints for a duality in section \ref{sect:all-plus-YM} below.
\end{itemize}

We saw that dual conformal symmetry is manifest when using the $x$-coordinates. We can use dual-conformal transformations, without loss of generality, to send $x_0$ to infinity,
\begin{align}
f_n (x_1,\ldots,x_n) := \lim_{x_0 \to \infty} (x_0^2)^4 \, F_n(x_1,\ldots,x_n;x_0) \,. 
\label{fnLimFn}
\end{align} 
This object contains the same amount of information as $F_n$, and dependence on $x_0$ can be easily restored in the given $f_n$. 
We also define the leading singularities of $F_n$ in this frame,
\begin{align}
r_n(x_1,\ldots,x_n) := \lim_{x_0 \to \infty} (x_0^2)^4  R_n(x_1,\ldots,x_n;x_0) \,. \label{rLimR}
\end{align}

The frame $x_0 \to \infty$ has the advantage that it is easy to connect the kinematics to familiar momentum space terminology.
This is seen as follows. In this frame, the leading singularities are Poincar\'{e}-invariant functions of $n$ points separated by light-like distances, i.e. they are invariant under space-time shifts 
and Lorentz transformations. Equivalently, we can say that they are functions of $n$ light-like momenta, cf. Eq.~(\ref{xp}). Since we have already fixed dual conformal symmetry, this is exactly the same kinematics as for a non-dual-conformal scattering amplitude. Indeed, we will see that the known results bear close resemblance to scattering amplitudes in non-supersymmetric Yang-Mills theory.

We extensively use Lorentz spinor notations for space-time coordinates $x_{\a\da} := x_\mu \sigma^{\mu}_{\a\da}$ and $x^{\da \a} := x^\mu \tilde{\sigma}^{\da\a}_\mu$, which are defined with the help of $2\times 2$ Pauli matrices, $\sigma^\mu = (1,\vec\sigma)$ and $\tilde{\sigma}^\mu= (1,-\vec\sigma)$. 
 The spinor indices are raised and lowered with the help of the antisymmetric tensors $\ep_{\a\b}$ and $\ep^{\da\db}$, such that $x^{\da\a} = \ep^{\da\db} \ep^{\a\b} x_{\b\db}$ and $x_{\a\da} = \ep_{\a\b} \ep_{\da\db} \, x^{\db \b}$. The product of matrices $x$ is the diagonal matrix $x_{\a\da} \, x^{\da\b} = x^2 \delta_{\a}^{\b}$. The difference of the space-time coordinates of two adjacent cusps is a light-like vector~\p{xp}, which factorizes in a pair of helicity spinors $\la,\tilde\la$,
\begin{align}
 p_{i}^{\da\a} = \la^{\a}_{i} \tilde\la^{\da}_{i} \,.
\end{align}
We also use a short-hand notation for the Lorentz invariant $\vev{a|x\, y|b} = \la_a^{\a} x_{\a\da} \, y^{\da\b} \la_{b\,\b}$ and $\vev{\la_a \la_b}= \la^\alpha_a \la_{b\,\alpha}$.
  
We prefer to use momentum twistor \cite{Hodges:2009hk,Mason:2009qx} variables instead of the space-time coordinates $x_1,\ldots,x_n,x_0$, since the former automatically resolve the null polygon constraints \p{xp}. We introduce $n$ momentum twistors $Z_1,\ldots,Z_n$ which are points in $\mathbb{P}^3$. A pair of twistors $Z_i$ and $Z_{i+1}$ defines a line in the twistor space $\mathbb{P}^3$, which corresponds to a space-time point $x_i \sim Z_{i} \wedge Z_{i+1}$,
\begin{align}
Z_i^{I} = ( \la^{\a}_i ,  x^{\da\beta}_i \la_{i\, \beta} ) \,, \qquad I=\alpha,\da  \,. \label{Twxla}
\end{align} 
The collection of $n$ pairwise intersecting lines in the twistor space corresponds to $n$ light-like separated space-time points $x_1,\ldots,x_n$.

In order to describe the Lagrangian space-time coordinate $x_{0}$, we introduce a pair of momentum twistors $Z_A$ and $Z_B$ specifying a line in $\mathbb{P}^3$,
\begin{align}
Z_{A}^I = (\la_A^{\a},  x_0^{\da\beta} \la_{A\,\beta}) \,,\quad
Z_{B}^I = (\la_{B}^{\a},  x_0^{\da\beta} \la_{B\,\beta})\,. \label{Z0}
\end{align}
The auxiliary spinors $\la_A$ and $\la_B$ are redundant in this formalism, since only the line $x_0 \sim Z_A \wedge Z_B$ has a geometric meaning. In the following we deal with functions of the line,  which are invariant under shifts $Z_A^I \to Z_A^I + \ep Z_{B}^I$ and $Z_{B}^I \to Z_{B}^I + \ep Z_{A}^I$. 

The four-bracket of momentum twistors provides a natural dual-conformal invariant. The twistor four-brackets with consecutive indices are space-time distances squared, i.e. 
\begin{align}
x_{ij}^2= \frac{\vev{i i+1 j j+1 } }{ \vev{i i+1}\vev{j j+1} }\,,
 \qquad x_{i0}^2=\frac{\vev{A B i i+1} }{\vev{i i+1}\vev{A B} }\,. \label{twbrTox2}
 \end{align}
The four-brackets with generic indices, e.g. $\vev{AB i j}$, contain both parity even and parity odd contributions. A twistor expression of the parity odd invariant \p{defeps012345} is provided in \p{epsTw}.

Because of the nonzero dual-conformal weight of $F_n$ at the Lagrangian point $x_0$, we cannot rewrite $F_n$ and its leading singularities $R_n$ (see eq.~\p{Fnl}) in $SU(2,2)$ invariant form solely in terms of twistor variables. In order to compensate nontrivial helicity weights, we have to introduce prefactors 
\begin{align}
\vev{AB} = \vev{\la_A,\la_{B}} = \ep_{\alpha \beta} \lambda_A^\alpha \lambda_B^\beta \,,
\end{align}
in the leading singularities.
In the following, we tacitly assume that the appropriate factor $\vev{AB}$ is taken into account when transforming between the twistor and space-time coordinates. Slightly abusing notations, 
we use the same letters to denote space-time and twistor objects, e.g. for leading singularities
\begin{align}
& R(x_1,\ldots,x_n;x_0) = \vev{AB}^4 \, R(Z_1,\ldots,Z_n;Z_A,Z_B) \,, \label{RRbar} 
\end{align}
hoping that the distinction will be obvious from the context, i.e. the variables used.
The momentum twistor version of the leading singularities $R$ is invariant under local rescalings $Z_{i} \to t Z_i$, $i=1,\ldots,n$, and scales as $R \to t^{-4} R$ at $Z_A \to t Z_A$ or $Z_{B} \to t Z_{B}$.

As we have already mentioned, the frame $x_0 \to \infty$ is very helpful. The twistor counterpart of the dual-conformal transformation $x_0 \to \infty$ is a $SU(2,2)$ transformation which sets $Z_A \wedge Z_{B}$ to be the infinity bi-twistor \cite{Mason:2009qx}. In other words, to reach that frame we substitute $Z_A^I = (0,\mu_{A}^{\da})$ and $Z_{B}^I = (0,\mu_{B}^{\da})$ where $[\mu_{A},\mu_{B}] = 1$, so that the only nonzero components of $Z_A \wedge Z_{B}$ are $\ep^{\da\db}$.

\subsection{$n$-point tree-level formula in different representations}

After having seen first examples for $n=4,5$ (eqs.~\p{Ftree4point} and \p{F50local}), here we present the tree-level results for $F_n$ for arbitrary number of particles. We present the formula in different ways, as we expect one representation to be more suitable for higher-loop generalization, and for making symmetries manifest.

\begin{figure}[t]
\begin{center}
\includegraphics[width=0.3\columnwidth]{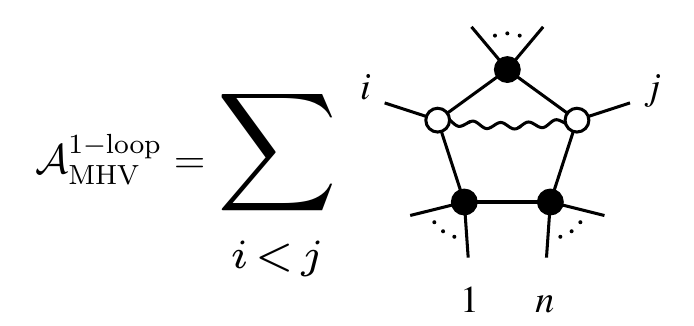}\quad
\includegraphics[width=0.4\columnwidth]{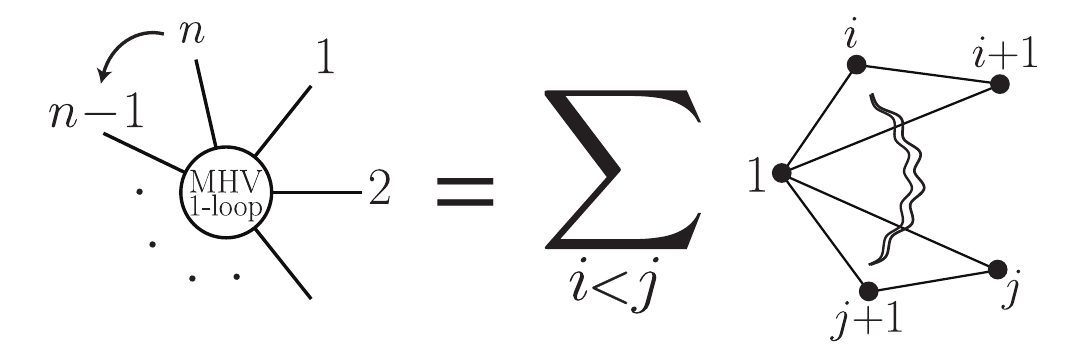}
    \caption{Left: `magic' pentagon representation \cite{ArkaniHamed:2010gh}. Right: `kermit' representation of one-loop MHV integrand \cite{Arkani-Hamed:2010zjl}. }
\label{fig:oneloop-kermit-pentagons}
 \end{center}
\end{figure}

We can benefit from the fact that the planar loop integrand for $F_n$ at $ L$ loops can be obtained from from $( L+1)$-loop amplitude integrands. Different forms of the one-loop $n$-point MHV integrand are available in the literature. We do not use the original box representation \cite{Bern:1994zx} since the parity odd integrand terms (that integrate to zero for the amplitude, but are relevant here) are not kept there. Using instead eq. (3.10) of \cite{ArkaniHamed:2010gh}, we immediately find
\begin{align}\label{oneloopfull1}
F_n^{(0)} = 
- \sum_{1<i<j<n}  \frac{\vev{ij1n} \vev{AB | (i-1 i i+1) \cap (j-1 j j+1)}}{\vev{AB i-1 i} \vev{AB i i+1} \vev{AB j-1 j} \vev{AB j j+1} \vev{AB1n}} \,.
\end{align} 
This is a representation in terms of pentagon integrands \cite{ArkaniHamed:2010gh}, see Fig.~\ref{fig:oneloop-kermit-pentagons}.
Here $(abc)\cap(def)$ stands for the line defined by the intersection of the planes $(abc)$ and $(def)$, given by
\begin{align}
\vev{xy(abc)\cap(def)} =& \vev{xabc}\vev{ydef}-\vev{yabc}\vev{xdef} \nonumber \\
 =& \vev{xyab} \vev{cdef} + \vev{xybc} \vev{adef} + \vev{xyca} \vev{bdef} \,. \label{twPlaneInters}
\end{align} 
Eq. (\ref{oneloopfull1}) has the expected dual conformal weights, and it has only local poles.
Although it is not obvious from the formula, the result does have the expected cyclic symmetry.
One may verify that the $n=4,5$ cases agree with Eqs. (\ref{Ftree4point}) 
and (\ref{F50local}), respectively, when converting them to momentum twistors. 

For our purposes, we prefer an alternative formula based on the BCFW representation of the one-loop MHV amplitude \cite{Arkani-Hamed:2010zjl} (dubbed `kermit' formula, see Fig.~\ref{fig:oneloop-kermit-pentagons}), which yields,
\begin{align}\label{kermit}
F_n^{(0)} = 
- \sum_{1<i<j<n} \frac{\langle AB| (1 i i+1)\cap (1 j j+1)\rangle^2}{\langle AB 1 i\rangle \langle AB i i+1\rangle \langle AB i+1 1\rangle\langle AB 1 j\rangle \langle AB j j+1\rangle \langle AB j+1 1\rangle}\,. 
\end{align}
In distinction to Eq. (\ref{oneloopfull1}), the representation (\ref{kermit}) contains spurious poles that cancel between the different terms. This is a normal feature of BCFW representations. 

Looking at eq.~(\ref{kermit}) it is natural to define the following five-point momentum twistor expression,
\begin{align}
B_{ijklm} =  & \frac{\vev{AB(m ij)\cap(jkl)}^2 }{\vev{AB ij}\vev{AB jk}\vev{AB kl}\vev{AB l j}\vev{AB jm }\vev{AB mi}} \,. \label{I5ptBos}
\end{align}
Furthermore, for certain `boundary terms' in eq. (\ref{kermit}), this formula simplifies, so that it is convenient to introduce the four-point twistor function
\begin{align}
B_{ijkl} = & \frac{ \vev{ijkl}^2}{\vev{AB ij}\vev{AB jk}\vev{AB kl}\vev{AB li}} \,. \label{I4ptBos} 
\end{align}
Let us rewrite eq. (\ref{kermit}) with explicit summation ranges, and writing out the boundary terms explicitly,
\begin{align}
F^{(0)}_{n}
=  \sum_{i=1}^{n-3} B_{i,i+1,n-1,n} - \sum_{i=1}^{n-4} B_{i,i+1,n-2,n-1,n} - \sum_{k=2}^{n-4}\sum_{i=1}^{n-k-3} B_{i,i+1,i+k,i+k+1,n}\,. \label{FnBosTree}
\end{align}
This formula will be a useful starting point for the following, in several respects. 
As we will see presently, the $B$'s are given by a Grassmannian formula, and can be thought of as leading singularities of $F_n$. While we do not have a general proof that all leading singularities take the form of the $B$'s, we will provide evidence for this at higher loops. Moreover, the form (\ref{FnBosTree}) will allow us to discover hidden conformal properties, and a connection to all-plus amplitudes in pure Yang-Mills theory.
Finally, the $B$ terms provide a good starting point for a supersymmetric extension \cite{CHTinpreparation}. 

Let us close this subsection by returning to the four- and five-cusp examples that we wrote earlier in $x$ notation.
Using the momentum twistor language, they are now expressed as follows,
\begin{align}
F_{4}^{(0)}  = B_{1234}  \,, \label{F40}
\end{align} 
and
\begin{align}
F_5^{(0)} =  B_{1245} + B_{2345} - B_{12345} \,. \label{F50}
\end{align} 
Note that individual terms in the linear combination (\ref{F50}) have spurious singularities which cancel out in the sum. In other words, $F_5^{(0)}$ contains only local poles, that is singularities when some $x_{ij}^2$ vanishes.

\begin{table}[t]
\begin{center}
\begin{tabular}{|l|c|c|c|c|c|c|c|}
\hline
$n$ &  4 & 5& 6 & 7 & 8 & 9 & 10 \\ \hline
\# $ B_{ijkl} $ & 1& 5& 15& 35 & 70 & 126 & 210 \\ \hline
\# $ B_{ijklm} $  & 0 & 1 & 6 & 21 & 56 & 126 & 252 \\ \hline
\# linearly independent & 1 & 6 & 20 & 50 & 105 & 196 & 336 \\ \hline
\end{tabular}\caption{Counting of the linearly independent twistor functions \p{I5ptBos} and \p{I4ptBos} in the $n$-cusp case assuming the cyclic ordering of the indices. Conjecturally, these rational functions give a basis of the leading singularities $R_n$ at any loop order.} \label{TabCountRes}
\end{center}
\end{table}

The benefits of using momentum twistors to describe the $B$'s becomes clear if we consider that when written in terms of the space-time variables $x_{ij}^2$, their expressions look rather complicated. See Appendix \ref{sec:Appendix:B-spacetime} for details.

It is interesting to enumerate the linearly independent functions. In the $n$-point case we take a collection of $\binom{n}{4}$ functions $B_{ijkl}$ with $1 \leq i < j < k < l \leq n$, and a collection of $\binom{n}{5}$ functions $B_{ijklm}$ with $1 \leq i < j < k < l < m \leq n$. Here we have imposed the cyclic ordering of indices which reflects the geometry of the Wilson loop contour. Both collections are linearly independent separately. Combining them together we find a number of linear relations counted in Tab.~\ref{TabCountRes} for $n\leq 10$. We observe that the number of linear-independent $n$-point dual-conformal invariants for the chosen cyclic ordering equals to   $(n-1)(n-2)^2(n-3)/12$. We checked this counting up to $n=15$.

\subsection{Conformal symmetry of the leading singularities}
\label{sec:subsectionconformalLS}

We now wish to reveal the conformal symmetry of the leading singularities $B_{ijkl}$ and $B_{ijklm}$ given in Eqs. \p{I5ptBos} and \p{I4ptBos}. In order to do so, we consider the frame $x_0 \to 
\infty$, which in twistor notation corresponds to $Z_A \wedge Z_{B} \to \ep^{\da\db}$. This will yield an expression written explicitly in the momentum space notation, using the usual $\lambda, \tilde{\lambda}$ spinor variables.
We denote 
\begin{align}
b =B |_{Z_A \wedge Z_{B} \to \ep^{\da\db}}\,. \label{bB}
\end{align}
Substituting each twistor  $Z_i$ by $(\la_i,  x_i \la_i)$, and evaluating the four-brackets as determinants, we arrive at
\begin{align}
b_{ijkl}  = & \frac{\bigl( \vev{ijkl} \bigr) ^2}{\vev{ij} \vev{jk} \vev{kl} \vev{l i}} \,, \label{Cfunc} \\
b_{ijklm}  = & \frac{ \bigl( \vev{ij} \vev{jklm} - \vev{j m} \vev{ijkl} \bigr)^2}{\vev{ij} \vev{jk} \vev{kl} \vev{kj} \vev{jm} \vev{mi}}  \,, \label{Bfunc}
\end{align}
with $1 \leq i < j < k < l \leq n$ and $1 \leq i < j < k < l < m \leq n$, respectively, where
\begin{align}\label{explicit-four-bracket-dualspace}
\vev{abcd} = & \vev{a |x_{a}  x_{b} | b} \vev{cd} - \vev{a | x_{a}  x_{c} | c} \vev{b} + \vev{a|x_{a} x_{d} | d} \vev{bc}  \notag\\
&  + \vev{b|x_{b}  x_{c} | c} \vev{ad} - \vev{b|x_{b} x_{d} | d} \vev{ac} + \vev{c|x_{c} x_{d} |d} \vev{ab} \,.
\end{align}
Recall the notation $\vev{a |x_{a}  x_{b} | b} = \lambda_a^{\alpha} (x_{a})_{\alpha \dot\alpha} (x_n)^{\dot \alpha \beta} \lambda^b{}_{\beta}$.

Since the functions $b_{ijkl}$ and $b_{ijklm}$ carry zero helicity weight with respect to each $\la_i$, they can be also expressed solely in terms of the coordinates $x_1,\ldots,x_n$.
The counting of linearly independent $b$'s is the same as that of $B$'s in Tab.~\ref{TabCountRes}. In the following, when discussing conformal symmetry and connection with non-supersymmetric amplitudes, it will be convenient to normalize the $b$'s with Parke-Taylor factors, which carry nontrivial helicity weights, 
\begin{align}
{\rm PT}_n = \frac{1}{\vev{12}\vev{23}\ldots\vev{n1}}\,. \label{PT}
\end{align} 

Let us consider as an example $n=5$. There are six linearly independent $b$'s, see Table~\ref{TabCountRes}. After normalizing by the Parke-Taylor prefactor ${\rm PT}_5$ \p{PT}, they take the following simple form in spinor helicity notation,
\begin{align}
& {\rm PT}_5\, b_{1234} = \frac{[23]^2}{\vev{45}\vev{51}\vev{14}} \;,\qquad 
{\rm PT}_5\, b_{2345} = \frac{[34]^2}{\vev{51}\vev{12}\vev{25}}\,, \notag\\
& {\rm PT}_5 \, b_{1345} = \frac{[45]^2}{\vev{12}\vev{23}\vev{31}} \;,\qquad  {\rm PT}_5 \, b_{1245}= \frac{[51]^2}{\vev{23}\vev{34}\vev{42}}\,,  \notag\\
& {\rm PT}_5 \, b_{1235} = \frac{[12]^2}{\vev{34}\vev{45}\vev{53}}  \;,\qquad
{\rm PT}_5 \, b_{12345} = \frac{[13]^2}{\vev{24}\vev{45}\vev{52}} \,. \label{5ptrres}
\end{align}
We are now in a position to find something remarkable. The above expressions are conformally invariant.
This can be seen as follows. The conformal boost operator in momentum space is given the the following second order generator \cite{Witten:2003nn},
\begin{align}
\mathbb{K}_{\a\da} =  \sum_{i=1}^n \frac{\pa^2}{\pa \la_{i}^{\alpha} \pa \tilde\la_{i}^{\dot\alpha}} \,. \label{KboostDef}
\end{align}
The quantities in Eq.~\p{5ptrres} are obviously conformal (for generic kinematics), i.e. they are annihilated by the operator in Eq.~(\ref{KboostDef}).

As we discussed earlier, the fact that $F_n$ has an underlying conformal symmetry is not completely surprising, as it is a feature of the amplitudes loop integrand, but this symmetry is obvious only when integrating over $x_0$, which is not the case for $F_n$. What we find here is that the symmetry is realized in a simple way in the dual conformal frame where $x_0 \to \infty$, which we can reach without loss of generality.
In section~\ref{sectionhigherorder}, we show that the conformal symmetry persists at arbitrary number of cusps $n$.

We wish to emphasize that the individual summands in eq. (\ref{F50local}), when multiplied by ${\rm PT}_5$, are {\it not} conformally invariant. 
The same is true for the individual terms of the local version of $F_n$ given in eq. (\ref{oneloopfull1}).
In contrast, the conformal invariance at $n$ points is true term by term for eqs.~(\ref{kermit}) and \p{FnBosTree}, when multiplied by ${\rm PT}_n$ and $x_0 \to \infty$.

We now discuss Grassmannian formulas for the leading singularities. The latter will allow us to prove the conformal invariance of the $b$'s for arbitrary $n$.

\subsection{Two Grassmannian formulas for the leading singularities}
\label{subsection:grassmannians}

In ref. 
\cite{Arkani-Hamed:2010zjl}, an interesting Grassmannian formula was presented for the leading singularities of $n$-particle one-loop MHV integrands, as a function of the external variables, and of the integration point.
Indeed, the different terms in the MHV formula (\ref{kermit}) above, and hence the $B$'s, correspond to certain residues in that Grassmannian integral.
In this section, we review this formula, and derive a second form that is useful in the frame $x_0 \to \infty$.

The relevant auxiliary geometry is the Grassmannian ${\rm Gr}(2,n+2)$ of 2-planes in $\mathbb{C}^{n+2}$. We specify a 2-plane by a pair of $\mathbb{C}^{n+2}$ vectors subjected to $GL(2)$ transformations and assemble their coordinates in $2\times (n+2)$ matrix 
\begin{align}
C = \begin{pmatrix}
c_A^1 & c_{B}^1 & c_{1}^1 & \dots & c_n^1 \\
c_A^2 & c_{B}^2 & c_{1}^2 & \dots & c_n^2
\end{pmatrix}  \,. \label{Cmat}
\end{align}
Following \cite{Arkani-Hamed:2010zjl}, we introduce contour integrals over the complex Grassmannian manifold ${\rm Gr}(2,n+2)$, 
\begin{align}
R_n(Z_1,\ldots,Z_n;Z_A,Z_B) =  \int D^{2n} C \frac{(AB)^2}{(12)(23) \ldots (n1)} \prod_{r=1}^2 \delta^{4} \left( \sum_{i \in \{ A,B,1,\ldots,n \} } c_i^r Z_i \right) . \label{RnBosGrInt}
\end{align}
where $(ab)$ denotes $2\times 2$ minors of the matrix $C$ \p{Cmat}. 
The notation for the integration measure is a shorthand for
\begin{align}
D^{2n} C = \frac{d^{2(n+2)}C}{{\rm vol}(GL(2))} \,. \label{DCmeasure0}
\end{align}
Explicitly, it is given by
\begin{align}
\frac{d^{2(n+2)}C}{{\rm vol}(GL(2))} = & \frac{1}{(n+2)!^2} \ep^{i_1 \ldots i_{n+2}} \ep^{j_1 \ldots j_{n+2}} (i_{n+1} i_{n+2}) (j_{n+1} j_{n+2}) \,\cdot \notag\\
& \times d c^1_{i_1} \wedge \ldots d c^1_{i_{n}} \wedge d c^2_{j_1} \wedge \ldots \wedge d c^2_{j_{n}}\,, \label{DCmeasure}
\end{align}
where the summation $i_1,\ldots,i_{n+2},j_1,\ldots,j_{n+2}$ is over $A,B,1,\ldots,n$. The measure $D^{2n} C$ is covariant under local $GL(2)$ transformations, namely $D^{2n} (g\, C) = (\det(g))^{n+2} D^{2n} C$
with $g \in GL(2)$. Taking into account $GL(2)$ transformations of the minors in \p{RnBosGrInt}, one can easily see that the integration is well-defined over the Grassmannian manifold.
We use the letter $R_n$ to indicate that we expect (\ref{RnBosGrInt}) to give the leading singularities of $F_n$.

The integral \p{RnBosGrInt} is a function of the twistor line $Z_{A}\wedge Z_{B}$, since any $GL(2)$ rotations of this pair of twistors can be compensated by a $GL(2)$ rotation of the first two columns of $C$ (see. eq.~\p{Cmat}). Indeed, the measure $D^{2n} C$ is covariant upon global $GL(n+2)$ rotations acting on $C$ \p{Cmat} from the right.  The compensating global $GL(2)$ transformation is embedded into $GL(n+2)$.

Let us count the dual conformal weights of the contour integral \p{RnBosGrInt}. We rescale one of the momentum twistors, $Z_i \to t Z_i$ with $i=1,\ldots,n$. The compensating rescaling of the $i$-th column of matrix $C$, i.e. $\vec{c}_i \to t^{-1}\vec{c}_i$, preserves $\vec{c}_i Z_i$ in the delta function argument in \p{RnBosGrInt}. The measure rescaling $D^{2n} C \to t^{-2} D^{2n} C$ is balanced by rescaling of the minors $(i-1 i)(i i+1) \to  t^{-2} (i-1 i)(i i+1) $. Thus, $R_n$ is invariant under the rescaling of the $Z_i$. On the other hand, rescaling one of $Z_a$ with $a=A,B$, i.e. $Z_a \to t Z_a$, the compensating transformation $\vec{c}_a \to t^{-1} \vec{c}_a$ results in $R_n \to t^{-4} R_n$. Indeed, the nontrivial weight comes from rescaling of the measure and $(AB)^2 \to t^{-2} (AB)^2$. This counting agrees with the discussion of the dual conformal weights of the leading singularities mentioned after eq.~\p{RRbar}.

The dual-conformal symmetry of $R_n$, namely covariance upon $GL(4)$ transformations of the momentum twistors, is also obvious in the contour integral representation \p{RnBosGrInt}.
In order to evaluate the contour integral, we can choose the coordinate patch 
\begin{align}
C \xrightarrow{GL(2)} 
\begin{pmatrix}
1 & 0 & c_1^1 & \dots & c_n^1 \\
0 & 1 & c_1^2 & \dots & c_n^2 
\end{pmatrix}\,, \label{Cpatch}
\end{align}
and then $D^{2n} C \to \prod_{r=1,2}\prod_{i=1}^n d c^{r}_{i} $. The bosonic delta functions freeze 8 integrations, and the remaining $(2n-8)$ complex integrations are done be taking residues.

We find it easier to study the residues in a different version of this formula that we derive presently.
Indeed, without loss of generality, let us study the contour integral representation \p{RnBosGrInt} in the dual conformal frame  $x_0 \to \infty$.
In twistor variables this amounts to
\begin{align}
r_n(Z_1,\ldots,Z_n) = R_n(Z_1,\ldots,Z_n;Z_A,Z_{B})|_{Z_A \wedge Z_{B} \to \ep^{\da\db}} \label{rnRn}
\end{align}
where we use notations $Z_i = (\la^{\a}_i,\mu^{
\da}_i)$ for the twistor components split according to the choice of the infinity bi-twistor.
Making this substitution, we obtain
\begin{align}
& r_n(Z_1,\ldots,Z_n) =
 \int  D^{2n} C \, \frac{(AB)^2 }{(12)(23) \ldots (n1)}  
\prod_{r=1}^{2} \delta^{2} \left( \sum_{i=1}^n c_i^r \la_i \right) \cdot  \notag\\ 
& \qquad\qquad\qquad\qquad\quad 
 \times  \delta \left( c^r_{A} + \sum_{i=1}^n c_i^r [\mu_i,\mu_{B}] \right) 
\delta \left( c^r_{B} + \sum_{i=1}^n c_i^r [\mu_{A},\mu_i] \right) , \label{eq3.1}
\end{align}
where we can choose a patch on the Grassmannian and integrate out $c^r_A,c^r_{B}$ with the help of the delta-constraints.
The remaining integrations are over $(2n-4)$-dimensional Grassmannian manifold ${\rm Gr}(2,n)$, with the measure $D^{2(n-2)}C$, see eq. \p{DCmeasure}. In other words, doing the dual-conformal transformation $x_0 \to \infty$ we contract the auxiliary geometry,
\begin{align}
{\rm Gr}(2,n+2) \to {\rm Gr}(2,n) \,.
\end{align} 
Finally, we arrive at the contour integral representation for the leading singularities in the frame $x_0 \to \infty$,
\begin{align}
r_n(Z_1,\ldots,Z_n) =  \int  D^{2 (n-2)} C
\frac{\left( \sum\limits_{i < j} (ij) [\mu_i \mu_j]\right)^2 }{(12)(23) \ldots (n1)}  \prod_{r=1,2}
 \delta^{2} \left( \sum_{i=1}^n c_i^r \la_i \right)  . \label{confGr}
\end{align}
The substitution $\mu^{\da}_i = x^{\da\a}_i \la_{i\,\alpha}$ converts the previous expression to space-time coordinates, 
\begin{align}
[\mu_i \mu_j] =  \vev{i|x_i  x_j |j} \,.
\end{align}
 It may seem that the resulting expression is not invariant under space-time shifts.
In fact, the integral in \p{confGr} is invariant under the
Poincar\'{e} part of the dual conformal algebra (see eqs. \p{dualconfAlgxlaGen0} and \p{dualconfAlgxlaGen}). Using the delta-function constraints from \p{confGr} we achieve the translation invariance 
\begin{align}\label{eq:shift:invariance}
\sum_{i < j} (ij) \vev{i|x_i x_j |j} \to \sum_{i < j} (ij) \vev{i|x_{ik} x_{kj} |j}\,,
\end{align} with any $k$. However, since we have chosen the frame $x_0 \to \infty$, the dual-conformal symmetry of \p{confGr} is broken.

The delta-function constraints in \p{confGr} freeze four integrations, and the remaining $2n-8$ integrations are done by the residue calculus. The residues are linear combinations of $b$'s \p{bB}, and counting of the linearly independent residues $r_n$ and the functions $b_{ijkl}$ and $b_{ijklm}$ is the same as in Table~\ref{TabCountRes}. 

Certain choices of contours in eqs. (\ref{RnBosGrInt}) and (\ref{confGr}) give the $B$'s and $b$'s, respectively. This follows from the connection to one-loop MHV scattering amplitudes, which correspond to $F_n$ at Born level. We find it plausible that at loop level these formulas govern the leading singularities of $F_n$. In the following we will present evidence for this hypothesis.

{The contour integral representation \p{confGr} of the $b$'s will be instrumental in proving their conformal symmetry and higher charge Yangian symmetries.
Moreover, the contour integral admits a BCFW-bridge-like decomposition in a sequence of twistor shifts,
thereby making a connection with the cell decomposition of the Grassmannian and plabic graphs \cite{Arkani-Hamed:2016byb}. Another remarkable feature of this concise expression for the conformal invariants is that for a particular integration contour it gives a non-supersymmetric loop amplitude, as will be discussed in Sect.~\ref{sect:all-plus-YM}.}

%
%
%

\section{Higher order symmetries and integrability}
\label{sectionhigherorder}

In this section, we study higher order symmetries of the Grassmannian formulas we proposed for $r_n$ and $R_n$, respectively. Recall that due to the close connection to scattering amplitudes, it is known that  integrating over $Z_{A}Z_{B}$ (for example, on a cycle giving leading singularities of the amplitudes), one obtains Yangian invariants. This means that $R_n$ (and hence also $r_n$, which corresponds to a particular dual conformal frame choice of $R_n$), should `know about' the Yangian symmetry: upon Yangian transformations acting on the external twistors, it should turn into a total derivative w.r.t. $Z_{A}Z_{B}$. On the one hand, it is unclear to us how useful this assertion is. On the other hand, in section \ref{sec:subsectionconformalLS} we have uncovered hints for a conformal invariance of $r_n$. This motivates us to look for a proof of this symmetry, and to look for higher-order constraints on both $R_n$ and $r_n$. We do so starting from the Grassmannian formulas.

We provide a proof of the conformal symmetry of the Grassmannian formula (\ref{confGr}) for $r_n$, and hence for the four-point and five-point functions $b$ given in eqs. (\ref{Cfunc}) and (\ref{Bfunc}). We also identify additional higher-order symmetries.

Furthermore, we carry out a similar analysis in the general frame, i.e. for arbitrary $x_0$. In this way, we can formulate analogues of the conformal symmetry relation, as well as of the higher-order equations found, directly for the rational functions $R_n$ and $B$. We find that this is closely related to differential equations found in ref. \cite{Ferro:2016zmx} in the context of the tree-level Amplituhedron.

%
%
%

\subsection{Proof of conformal and higher-order symmetries of the leading singularities $r_n$}
\label{subsectionproof1}

\subsubsection{Yangian generators}

The dual-conformal algebra generators have a very simple representation when written in momentum twistors $Z^I=(\la^\alpha,\mu^{\da})$, namely\footnote{We could have chosen to supplement $su(2,2)\approx sl(4)$ with the central element $H = Z^L \frac{\pa}{ \pa Z^L}$ and consider $gl(4)$ generators $\mathfrak{J}^{I}{}_{J} = Z^I \frac{\pa}{\pa Z^J}$. Indeed, each local instance of the central element annihilates the leading singularities $H_i \, r_n = 0$, $i=1,\ldots,n$, and commutes with other operators.}
\begin{align}
\mathfrak{J}^{I}{}_{J} = Z^I \frac{\pa}{\pa Z^J} - \frac{\delta^I_J}{4}  Z^L \frac{\pa}{\pa Z^L}  \,,
\label{dualConfJmat}
\end{align}
where $I=\a,\da$ and $J=\b,\db$.
Indeed, in this formula the $4\times 4$ matrix $\mathfrak{J}^{I}{}_{J}$ comprises the dual conformal generators $D,P,M,\bar M, K$ (see Appendix~\ref{AllDualConf}), 
\begin{align}
\mathfrak{J}^{I}{}_{J} = \begin{pmatrix}
M^{\a}{}_{\b} - \frac{\delta^{\a}_{\b}}{2} D & P^{\a}{}_{\db} \\
K^{\da}{}_{\b} & \bar M^{\da}{}_{\db}+\frac{\delta^{\da}_{\db}}{2} D 
\end{pmatrix} \,. \label{dualConfJmat2}
\end{align}
In the following we wish to act on the external momentum twistors $Z_{i}$, with $i=1,\ldots,n$. We denote by $\mathfrak{J}_i$ the local generator acting on the $i$-th coordinate, and we represent the dual-conformal generators as sums of the local ones
\begin{align}
\mathfrak{J}^{I}{}_{J}  =& \sum_{i=1}^n \left(\, \mathfrak{J}_i \, \right)^{I}{}_{J}  \;, \nonumber\\ \left(\, \mathfrak{J}_i  \, \right)^{I}{}_{J} =& Z_{i}^I \frac{\pa}{\pa Z_{i}^J} - \frac{\delta^I_J}{4}  Z_i^L \frac{\pa}{\pa Z_i^L} \,. \label{JgenSum}
\end{align}
Let us start by discussing the obvious symmetries. The leading singularities $r_n$  given in eq. (\ref{confGr}) are Poincar\'{e} invariant,
\begin{align}
P_{\a\db}\, r_n = M_{\a\b} \, r_n = \bar{M}_{\da\db} \, r_n = 0 \,, \qquad 
D\, r_n = 2 \, r_n \,, \label{Poinc}
\end{align}
but due to our frame choice $x_0 \to\infty$ the dual conformal invariance is broken,
\begin{align}
K_{\da\b} \, r_n \neq 0 \,. \label{noK}
\end{align}

In order to study higher symmetries of the leading singularities $r_n$, we introduce an infinite-dimensional Yangian algebra which extends the dual-conformal algebra. The Yangian of $su(2,2)$ is a Hopf algebra with an infinite tower of generators grouped into levels. The zeroth level generators 
$\mathfrak{J}^{(0)}  := \mathfrak{J} $ are those of $su(2,2)$, and the level-one generators
$\mathfrak{J}^{(1)} :=  \widehat{\mathfrak{J}}$ transform in the adjoint representation of $su(2,2)$,
\begin{align}
\left[ \mathfrak{J}^{I}{}_{J} , \widehat{\mathfrak{J}}^{K}{}_{L} \right] = \delta^K_J \,\widehat{\mathfrak{J}}^{I}{}_{L} - \delta^I_L \, \widehat{\mathfrak{J}}^{K}{}_{J} \,. \label{JJhatComm}
\end{align}
Higher-level generators are defined as repeated commutators of the level-one generators, subjected to the Serre relations~\cite{Drinfeld:1985rx}. 
 
The coproduct structure of the Hopf algebra suggests that we represent the level-one generators $\widehat{\mathfrak{J}}$ of the Yangian as sums of bi-local terms formed from the local generators $\mathfrak{J}_i $,  
\begin{align}
\widehat{\mathfrak{J}}^{I}{}_{J} = \sum_{1\leq i<j\leq n} 
\left( \mathfrak{J}_i \right)^{I}{}_{K} 
\left( \mathfrak{J}_j \right)^{K}{}_{J} \,. \label{JJgenSum}
\end{align}

\subsubsection{Spin-chain picture of the Yangian symmetry}
\label{sec:symm_rn}

As mentioned in the Introduction, here we use constructions borrowed from the quantum inverse scattering method and integrable spin chains to formulate the Yangian symmetry. 
 
Let us define the {\it Lax operator} ${\rm L}(u)$, which is a $4\times 4$ matrix with first order differential operator entries (dual-conformal generators $\mathfrak{J}^I{}_J$ \p{dualConfJmat}), 
\begin{align}
\left[ {\rm L}(u) \right]^{I}{}_J = u \,\delta^I_J + Z^I \frac{\pa}{\pa Z^J} - \frac{\delta^{I}_J}{4} Z^K \frac{\pa}{\pa Z^K} \,, \label{Lax}
\end{align}
and $u \in \mathbb{C}$ is called the spectral parameter. Let us denote by ${\cal H}_i$ the space of  homogeneous functions of $Z_i^I$ with homogeneity degree zero, i.e. functions which are invariant under rescaling $G(Z_i^I) = G(t Z_i^I)$.  Then ${\rm L}_i(u)$ acts on ${\cal H}_i \otimes \mathbb{C}^4$.

We associate a quantum spin chain to the leading singularities $r_n$, such that each $i$-th site of the spin chain carrying representation space $\mathcal{H}_i$ corresponds to the $i$-th edge of the Wilson loop. We assign a Lax operator to each site of the spin chain and we form the matrix product of $n$ Lax operators, which is called the {\it monodromy matrix},
\begin{align}
\left[ {\rm T}(u) \right]^I{}_J:= \left[{\rm L}_1(u)\right]^{I}{}_{K_1} \left[{\rm L}_2(u) \right]^{K_1}{}_{K_2} \ldots \left[{\rm L}_n(u)\right]^{K_{n-1}}{}_J \,. \label{TmatMonodr}
\end{align}
The monodromy matrix ${\rm T}(u)$ acts in the space
\begin{align}
{\cal H}_1 \otimes {\cal H}_2 \otimes \ldots \otimes {\cal H}_n \otimes \mathbb{C}^4 \,.
\end{align}
Expanding the monodromy matrix in the spectral parameter $u$ we recover the Yangian generators in the evaluation representation \cite{Molev:1994rs}: the dual-conformal generators $\mathfrak{J} := \mathfrak{J}^{(0)}$ \p{dualConfJmat} at the zeroth level, first level generators $\widehat{\mathfrak{J}} := \mathfrak{J}^{(1)}$ \p{JJgenSum} given by the sum of bi-local terms, as well as higher level generators $\mathfrak{J}^{(2)},\ldots, \mathfrak{J}^{(n-1)}$,
\begin{align}
\left[ {\rm T}(u)\right]^{I}{}_J = u^n \delta^I_J + u^{n-1} \left[ \mathfrak{J}^{(0)}\right]^{I}{}_J + u^{n-2} \left[\mathfrak{J}^{(1)}\right]^{I}{}_J  + \sum_{k=2}^{n-1} u^{n-k-1} \left[  \mathfrak{J}^{(k)} \right]^{I}{}_J \,, \label{Texpand}
\end{align}
with $k$-th level generator being the sum of ordered products of $k+1$ local generators \p{JgenSum},   
\begin{align}
\left[  \mathfrak{J}^{(k)} \right]^{I}{}_J = \sum_{1 \leq i_1 < i_2 < \ldots < i_{k+1} \leq n} 
\left( \mathfrak{J}_{i_1} \right)^{I}{}_{L_1} 
\left( \mathfrak{J}_{i_2} \right)^{L_1}{}_{L_2} \ldots \left( \mathfrak{J}_{i_{k+1}} \right)^{L_{k}}{}_{J} 
\,. \label{Jfull}
\end{align}
In the following we denote $\mathfrak{J}^{(0)}$ and $\mathfrak{J}^{(1)}$ as $\mathfrak{J}$ and $\widehat{\mathfrak{J}}$, respectively, to simplify notations.
In eq. (\ref{Jfull}) a specific choice of origin of the sum was made. Below we also discuss the cyclically related choices.

Let us note that in the evaluation representation \p{TmatMonodr} of the Yangian induced by \p{Lax} the infinite tower of Yangian generators in truncated at the level $n$. 
Studying the Yangian symmetry, we work with the monodromy matrix \p{TmatMonodr}, which encompasses all Yangian generators in the chosen representation. Instead of acting with individual Yangian generators $\mathfrak{J}^{(k)}$ on $r_n$, we would like to calculate the action of the monodromy matrix $\mathrm{T}(u)$ on $ r_n$. Moreover, we would like to reduce the evaluation of $\mathrm{T}(u)\, r_n$ to manipulations with local objects, those which act nontrivially on a single or a pair of spin chain sites. By definition \p{TmatMonodr}, the monodromy matrix is a product of local blocks. Therefore, we will also need a decomposition of $r_n$ into local blocks.

We define an ${\rm R}$-operator which acts on a pair of the spin chain sites ${\cal H}_i \otimes {\cal H}_j$ and transforms a test function $G$ as follows,
\begin{align}
\left[ {\rm R}_{ij} \, G \right](Z_i,Z_j) := \int \frac{dt}{t} G(Z_i + t Z_j,Z_j) \,. \label{Rop}
\end{align}
We imply that the integration contour lies in the complex plane and it is closed, but we do not fix it. Let us note that the $\mathrm{R}$-operator preserves homogeneity, hence it does not map out of ${\cal H}_i \otimes {\cal H}_j$. 
The $\mathrm{R}$-operator \p{Rop} commutes with the matrix product of the pair of Lax operators \p{Lax},
\begin{align}
& {\rm R}_{ij} \, {\rm L}_i(u)\, {\rm L}_j(u)\, = {\rm L}_i(u)\, {\rm L}_j(u)\,{\rm R}_{ij} \,, \\
& {\rm R}_{ji} \, {\rm L}_i(u)\, {\rm L}_j(u)\, = {\rm L}_i(u)\, {\rm L}_j(u)\,{\rm R}_{ji} \,,
\end{align} 
as can be easily verified. Let us stress that the $\mathrm{R}$-operator acts trivially in the matrix space of the previous equations.
Thus, the $\mathrm{R}$-operator acting in adjacent spin chain sites commutes with the monodromy matrix \p{TmatMonodr},
\begin{align}
{\rm R}_{i\, i+1}\, {\rm T}(u) = 
{\rm T}(u) \, {\rm R}_{i\, i+1} \;,\qquad
{\rm R}_{i+1\, i}\, {\rm T}(u) = 
{\rm T}(u) \, {\rm R}_{i+1\, i} 
\;,\quad i=1,\ldots,n-1 \,.  \label{TRcomm}
\end{align}

Now we use a corollary proven in section \ref{subsection:corollaryRoperatorsGrassmannian}.
There we show that the contour integral \p{confGr} over the Grassmannian ${\rm Gr}(2,n)$
can be rewritten as a product of $2(n-2)$ $\mathrm{R}$-operators acting on a function $\ket{\Omega_n}$, 
\begin{align}
r_n(Z_1,\ldots,Z_n) = {\rm R}_{32}\,{\rm R}_{43}\, \ldots {\rm R}_{n \, n-1}\, {\rm R}_{n-2\,n-1}\, \ldots {\rm R}_{23}\,{\rm R}_{12}\,\ket{\Omega_n} \,. \label{goodRseq}
\end{align} 
Each of the $\mathrm{R}$-operators from the product in \p{goodRseq} introduces one-dimensional integration.
We call $\ket{\Omega_n}$ a pseudo-vacuum state of the $n$-site spin chain,
\begin{align}
\ket{\Omega_n} := \delta^2(\la_1)\delta^2(\la_n) [\mu_1\mu_n]^2 \,. \label{vacuum}
\end{align} 
One can easily see that the distribution $\ket{\Omega_n}$ is of homogeneity degree zero, i.e. it is an element of the space ${\cal H}_1 \otimes \ldots \otimes {\cal H}_n$.

The employed notations are very suggestive. The pseudo-vacuum $\ket{\Omega_n}$ of the spin chain represents a zero-dimensional cell of the Grassmannian ${\rm Gr}(2,n)$. Each `excitation' over the pseudo-vacuum introduced by the $\mathrm{R}$-operator is a BCFW bridge \cite{Arkani-Hamed:2016byb}, which maps to a Grassmannian cell of higher dimension. There are $2(n-2)$ `excitations' in \p{goodRseq} that correspond to the top cell of the Grassmannian.

Since the monodromy matrix $\mathrm{T}(u)$ commutes with individual $\mathrm{R}$-operators \p{TRcomm}, then it also commutes with their products in \p{goodRseq}. Thus, in order to calculate $\mathrm{T}(u)\, r_n$ it remains to find how the $4\times 4$  monodromy matrix \p{TmatMonodr}  acts on the pseudo-vacuum $\ket{\Omega_n}$ \p{vacuum}. We find that it takes a block-triangular form
\begin{align}
{\rm T}(u) \, \ket{\Omega_n} = u^{n-2} \begin{pmatrix}
 u^2-2 u +1 & 0 \\
0 & u^2+2u-1
\end{pmatrix} \ket{\Omega_n} + u^{n-2} \begin{pmatrix}
0 & 0 \\
f+ u g &  h
\end{pmatrix} , \label{eq:TOmega}
\end{align}
where dependence of the r.h.s. on the spectral parameter $u$ is explicit; $f,g,h$ are $2
\times 2$ matrices which depend on twistor variables $Z_1$ and $Z_n$, and matrix $h$ is traceless, $\tr(h) = 0$. The explicit expressions for $f,g,h$ will be irrelevant in what follows.
Eq. (\ref{eq:TOmega}) expression is easy to work out, since by acting on the pseudo-vacuum successively with each of the constituent Lax operators $\mathrm{L}(u)$ given in eq. \p{Lax}, we obtain either diagonal or block-triangular matrices. 

Finally, eqs. \p{eq:TOmega}  and \p{goodRseq} enable us to find how the monodromy matrix acts on the leading singularities $r_n$, namely
\begin{align}
{\rm T}(u) \, r_n = u^{n-2} \begin{pmatrix}
 u^2-2 u +1 & 0 \\
0 & u^2+2u-1
\end{pmatrix} r_n + u^{n-2} \begin{pmatrix}
0 & 0 \\
{\cal O}(u) &  {\cal O}(u^0)
\end{pmatrix}  ,  \label{eq:Trn}
\end{align}
where the last matrix term of the r.h.s. is traceless.

These equations embody our statement on the partial Yangian symmetry of $r_n$. They contain the complete collection of the Yangian invariances satisfied by $r_n$ with Yangian generators of the form \p{Jfull}.

Expanding eq.~\p{eq:Trn} in the spectral parameter $u$ according to \p{Texpand}, we find how the Yangian generators of all levels \p{Jfull} act on $r_n$. At order $u^{n-1}$ in \p{eq:Trn}, we recover Poincar\'{e} invariance \p{Poinc} with $\mathfrak{J} = \mathfrak{J}^{(0)}$,
\begin{align}
\mathfrak{J}^{\a}{}_{\b}\, r_n = -2 \delta^{\a}_{\b} \, r_n \; , \quad  
\mathfrak{J}^{\a}{}_{\db}\, r_n =0 \;, \quad \mathfrak{J}^{\da}{}_{\db}\, r_n = 2 \delta^{\da}_{\db} r_n \,. \label{eq:main-Yangian-rn1}
\end{align} 
At order $u^{n-2}$, we find that a half of the level-one generators $\mathfrak{J}^{(1)}=\widehat{\mathfrak{J}}$ provide invariances,
\begin{align}
\widehat{\mathfrak{J}}^{\a}{}_{\b}\, r_n = \delta^{\a}_{\b} \, r_n \; , \quad  
\widehat{\mathfrak{J}}^{\a}{}_{\db}\, r_n =0 \;, \quad \widehat{\mathfrak{J}}^{\da}{}_{\da}\, r_n = -2 r_n \,. \label{eq:main-Yangian-rn2}
\end{align} 
However, the level one-generators $\widehat{\mathfrak{J}}^{\da}{}_{\b}$ and the traceless part of $\widehat{\mathfrak{J}}^{\da}{}_{\db}$ are not symmetries of $r_n$. In other words, this part of the Yangian symmetry is broken.

We observe that there are no terms $u^{n-1-k}$ with $k=2,\ldots,n-1$ on the RHS of \p{eq:Trn}. Consequently, all Yangian generators at the level two and higher provide invariance relations, 
\begin{align}
\left(\mathfrak{J}^{(k)}\right)^{I}{}_{J}\, r_n =0 \,, \qquad k = 2 ,\ldots,n-1\,. \label{eq:main-Yangian-rn3}
\end{align}

In a similar fashion, we can prove additional relations.
Let us introduce the second order differential operators 
\begin{align}
D^{(i)}_{I} := \frac{\pa}{\pa Z_i^J} \left( \sum_{j=1}^n Z_j^J \frac{\pa}{\pa Z_j^I}\right) - 2 \frac{\pa}{\pa Z_i^I} \;,\qquad i = 1,\ldots,n \,.
\end{align} 
They will play the role of the monodromy matrix from the previous discussion. Indeed, we can check by explicit calculation that the
operators $D^{(i)}_I$ commute with any $\mathrm{R}$-operator \p{Rop} and annihilate the pseudo-vacuum $\ket{\Omega_n}$ \p{vacuum},
\begin{align}
& D^{(i)}_I \,  \ket{\Omega_n} = 0 \, , \notag\\
& \left[ {\rm R}_{kl} \, ,\,  D^{(i)}_I \right] = 0 \,,\qquad i,k,l = 1,\ldots, n\,.
\end{align}
Thus, we can pull $D^{(i)}_I$ through the product of $\mathrm{R}$-operators in representation \p{goodRseq} and find that  it annihilates the leading singularities $r_n$,
\begin{align}\label{eq:Dbonus-rn}
D^{(i)}_{I}  \, r_n  
= 0 \; , \qquad i = 1,\ldots, n\,.
\end{align}
These relations can also be formulated in terms of the dual-conformal algebra generators \p{JgenSum} as follows,
\begin{align}
\left( \, \mathfrak{J}_i \, \right)^{K}{}_{J} \, \mathfrak{J}^{J}{}_{I} \, r_n = 2 \, \left( 
\, \mathfrak{J}_i \, \right)^{K}{}_{I} \, r_n \;, \qquad i=1,\ldots, n \,.
\label{bonusDiffEq}
\end{align}
In Appendix.~\ref{sec:cyclYangK}, we rewrite eq.~\p{bonusDiffEq} in terms of the dual-conformal generators acting on the space-time variables. This enable us to see in more details how the frame choice $x_0 \to \infty$ for $r_n$ breaks the dual-conformal symmetry.

Another way to interpret eqs. \p{eq:Dbonus-rn} and \p{bonusDiffEq} is to consider the cyclic shift transformations of the Yangian. The ordering of the Wilson loop edges is irrelevant for the level-zero generators $\mathfrak{J}^{(0)}$. However, the level-one and higher-level generators of the Yangian \p{Jfull} employ the cyclic ordering with the specific choice of the sum origin. At the same time, the contour integral representing the leading singularities $r_n$ is invariant under cyclic shifts $Z_i \to Z_{i+1}$, see \p{confGr}. Thus, we can formulate the Yangian symmetries of $r_n$ using a set of cyclically shifted generators. Then we sum in \p{Jfull} over cyclically shifted labels. We provide the transformation properties of the Yangian generators upon the cyclic shift in Appendix~\ref{AppCycl}. Let us consider the simplest nontrivial cyclic-shift transformation. Indeed, along with the level-one generator $\mathfrak{J}^{(1)} = \widehat{\mathfrak{J}}$ \p{Jfull} we introduce the level-one generator $\widehat{\mathfrak{J}}'$ which employs the ordering $n,1,2,\ldots,n-1$. The level-one Yangian invariances from \p{eq:main-Yangian-rn2} hold with the cyclically shifted generator $\widehat{\mathfrak{J}}'$ as well. For example, the following difference of the generators annihilates $r_n$, 
\begin{align}
\left[ \widehat{\mathfrak{J}}^{\alpha}{}_{\beta} - \widehat{\mathfrak{J}}'{}^{\alpha}{}_{\beta}  \right] r_n = 0 \,. \label{eq:JminusJ}
\end{align}
Evaluating the difference of generators on the space of homogeneous functions of degree zero, i.e. $(\mathfrak{J}_n)^{C}{}_{C}=0$,
\begin{align}
\widehat{\mathfrak{J}}^{A}{}_{B} - \widehat{\mathfrak{J}}'{}^{A}{}_{B} = 4 (\mathfrak{J}_n)^{A}{}_{B} + \mathfrak{J}^{A}{}_{C} (\mathfrak{J}_n)^{C}{}_{B} - (\mathfrak{J}_n)^{A}{}_{C} \mathfrak{J}^{C}{}_{B} \,,
\end{align}
and choosing the indices $A,B$ as in \p{eq:JminusJ}, we conclude that \p{eq:JminusJ} is equivalent to \p{eq:Dbonus-rn} with $i=n$. Choosing other cyclic shifts we can reproduce all $n$ relations \p{eq:Dbonus-rn}. Let us also note that the cyclic consistency of the remaining Yangian invariances \p{eq:main-Yangian-rn2} does not provide new relations. 

Despite of the fact that a half of the level-one Yangian generators $\mathfrak{J}^{(1)} = \widehat{\mathfrak{J}}$ is broken, we can formulate several invariance relations for $r_n$ involving the broken generators and the dual-conformal generators, e.g.
\begin{align}
\mathfrak{J}^{\a}{}_{\dot\gamma} \, \widehat{\mathfrak{J}}^{\dot\gamma}{}_{\dot\beta} \, r_n & = 0 \;,\quad &
\mathfrak{J}^{\da}{}_{\dot\gamma} \, \widehat{\mathfrak{J}}^{\dot\gamma}{}_{\dot\beta} \, r_n & = 2 \delta^{\dot\alpha}_{\dot\beta} \, r_n + 4 \widehat{\mathfrak{J}}^{\dot\alpha}{}_{\dot\beta} \, r_n \,, \label{JJhat1}
\\
\mathfrak{J}^{\a}{}_{\dot\gamma} \, \widehat{\mathfrak{J}}^{\dot\gamma}{}_{\beta} \, r_n & = 4 \delta^{\a}_{\b} \, r_n \;,\quad &
\mathfrak{J}^{\da}{}_{\dot\gamma} \, \widehat{\mathfrak{J}}^{\dot\gamma}{}_{\beta} \, r_n & = 4 \widehat{\mathfrak{J}}^{\da}{}_{\beta}  \, r_n \,. \label{JJhat2}
\end{align}
In other words, even though the Yangian variations $\widehat{\mathfrak{J}}^{\dot\alpha}{}_{\beta}$ and $\widehat{\mathfrak{J}}^{\dot\alpha}{}_{\dot\beta}$ of $r_n$ are not covariant, contracting the variations with the dual-conformal generators we find symmetry relations for $r_n$. We deduce them from the commutation relations among $\mathfrak{J}$ and $\widehat{\mathfrak{J}}$ \p{JJhatComm} and the invariance relations \p{eq:main-Yangian-rn2}. Relations (\ref{JJhat1}) and (\ref{JJhat2}) are closely related to differential equations found in \cite{Ferro:2016zmx}, as will be discussed in section \ref{subsectionproof4}.

%
%
\subsubsection{Summary and discussion of symmetry relations}

Using integrability methods, we have proven the Yangian invariance relations (\ref{eq:main-Yangian-rn2}), (\ref{eq:main-Yangian-rn3}) as well as the bonus relations (\ref{bonusDiffEq}), for the Grassmannian formula (\ref{confGr}) for $r_n$. Let us now discuss these results.

We see that eqs.  (\ref{eq:main-Yangian-rn2}) involve half of the level-one Yangian generators. This may be due to, at least in part, to our choice of fixing the dual conformal frame $x_0 \to \infty$.

Among the level-one Yangian invariances in eq. (\ref{eq:main-Yangian-rn2}), the one with $\widehat{\mathfrak{J}}^{\alpha}{}_{\dot\beta}$
has a transparent meaning. It is a conformal invariance in momentum space.
 In order to see this, let us make contact with the conformal boost generator $\mathbb{K}_{\a\db}$ given in Eq.~\p{KboostDef}, which is written in terms of the spinor-helicity variables $(\lambda, \tilde{\lambda})$, see Appendix~\ref{AppKvarChange} and \cite{Drummond:2010qh}. There it is shown that
\begin{align}  \label{Kinv}
  \widehat{\mathfrak{J}}
 {}^{\a}
 {}_{\db}
  \, r_n (Z_1,\ldots,Z_n)
  = - {\vev{12}\vev{23} \ldots \vev{n1}} \,   \mathbb{K}^{\a}{}_{\db} \, \frac{r_n(p_1,\ldots,p_n)}{\vev{12}\vev{23}\ldots\vev{n1}} \,.
\end{align}
In view of Eq.~\p{eq:main-Yangian-rn1} this implies
\begin{align}
\mathbb{K}^{\a}{}_{\db} \,  \frac{r_n(p_1,\ldots,p_n)}{\vev{12}\vev{23}\ldots\vev{n1}}  = 0 \,. \label{Kinv2}
\end{align}
Therefore we have proven the conformal invariance of the leading singularities $r_n$. This is one of the main results of this section.

%
%

\subsection{Formulation of higher-order symmetries of $r_n$ in momentum space}\label{App:subsection-momentum-symmetries}

Above we constructed higher charge symmetries \p{Jfull} of $r_n$ out of the local dual-conformal generators. Then we identified the conformal boost symmetry of $r_n$ \p{Kinv2} among the level-one Yangian generators. We could wonder whether it is possible to construct higher charge symmetries out of the conformal generators, which act on the momentum variables. In this section we briefly comment on this possibility.

Indeed, let us assemble the conformal algebra generators written in spinor helicity variables \cite{Witten:2003nn} in the $4\times 4$ matrix
\begin{align}
\mathbb{J}^I{}_{J} = \begin{pmatrix}
\la^{\a} \frac{\pa}{\pa \la^\b} & - \la^\a \tilde\la^\db \\[0.4em]
\frac{\pa^2}{\pa \tilde \la^{\da} \pa \la^\b} & - \frac{\pa}{\pa \tilde\la^{\da}} \tilde\la^{\db}
\end{pmatrix} = \begin{pmatrix}
\la^{\a} \\ \frac{\pa}{\pa \tilde\la^{\da}}
\end{pmatrix} \otimes \begin{pmatrix}
\frac{\pa}{\pa \la^\b} & -\tilde\la^{\db}
\end{pmatrix} \label{Jconf}
\end{align}
where $I = (\a,\da)$ and $J = (\b,\db)$, and for the sake of brevity we omitted index $i=1,2,\ldots,n$ which labels the particles. Summing local generators $\mathbb{J}_i$ over $i$, we obtain the conformal symmetries of $r_n$. In particular $I=\da$ and $J=\b$ corresponds to the conformal boost \p{KboostDef},
\begin{align}
\mathbb{J}^{(0)}_{\da\b} :=
\sum_{i=1}^n \left( \mathbb{J}_{i} \right)_{\da\b} = 
 \mathbb{K}_{\da\b} \,.
\end{align}

In analogy with \p{Jfull}, we define higher charge non-local generators out of the local conformal generators \p{Jconf}. The level-$k$ generators are the following sums of the ordered products of \p{Jconf},
\begin{align}
\left[  \mathbb{J}^{(k)} \right]^{I}{}_J := \sum_{1 \leq i_1 < i_2 < \ldots < i_{k+1} \leq n} 
\left( \mathbb{J}_{i_1} \right)^{I}{}_{L_1} 
\left( \mathbb{J}_{i_2} \right)^{L_1}{}_{L_2} \ldots \left( \mathbb{J}_{i_{k+1}} \right)^{L_{k}}{}_{J} \,. \label{JconfFull}
\end{align}

It would be interesting to systematically study the properties of $r_n$ under these generators, using methods similar to the last subsection. 
For the sake of this paper however, we limit ourselves to `experiment' with the differential operators (\ref{JconfFull}), and find, conjecturally, a set of equations that the $r_n$ satisfy.

Let us consider $I=\da$ and $J=\b$ in \p{JconfFull}, which is given explicitly by the following differential operator in terms of helicity spinors,
\begin{align}
& \mathbb{J}^{(k)}_{\da\b} = \sum_{1 \leq i_1 < i_2 < \ldots < i_{k+1} \leq n} \frac{\pa}{\pa \tilde\la_{i_1}^{\da}} \, Q_{i_1 i_2} \ldots Q_{i_{k} i_{k+1}} \frac{\pa}{\pa \la_{i_{k+1}}^\b} \,,\notag\\
& Q_{ij} := \la_{j}^\gamma \frac{\pa}{\pa \la_{i}^\gamma} - \tilde\la_{i}^{\dot\gamma} \frac{\pa}{\pa \tilde\la_{j}^{\dot\gamma}} \,.
\end{align}
These are level-$k$ counterparts of the conformal boost generator.
We conjecture that the following equations hold,
\begin{align}
\mathbb{J}^{(k)}_{\da\b} \; \frac{r_n(p_1,\ldots,p_n)}{\vev{12}\vev{23} \ldots \vev{n1}}\Biggr|_{\substack{\tilde\la_1 \to \frac{1}{\vev{n1}}(p_2 + \ldots +p_{n-1})\ket{n} \\ \tilde\la_n \to \frac{1}{\vev{1n}}(p_2 + \ldots +p_{n-1})\ket{1} }} = 0 \;, \qquad k=0,1,\ldots,n \,. \label{Jmom}
\end{align}
In this formula it is understood that $\tilde\la_1$ and $\tilde\la_n$ are eliminated with the help of momentum conservation prior to differentiations (this choice is related to the ordering of local generators in \p{JconfFull}). 

In order to
check eq. \p{Jmom}, we choose $r_n$ to be the four-point $b_{ijkl}$ \p{Cfunc} and five-point $b_{ijklm}$ \p{Bfunc} invariants  written in spinor helicity variables $(\la,\tilde\la)$ with  $\tilde\la_1$ and $\tilde\la_n$ eliminated as specified above. Thus, the four-point and five-point invariants nontrivially depend on all $n$ momenta. Following this approach, we checked eq. \p{Jmom} up to $n=15$. 

It would be interesting to find a proof of eq. \p{Jmom} for any $n$ which employs the integrability constructions above.  As mentioned before, in order to achieve this goal one would need a momentum space analogue of the contour integral representation \p{confGr}, e.g. similar to the formulas in \cite{ArkaniHamed:2009dn} for scattering amplitudes.

%
%
%

\subsection{Higher-order symmetries of $R_n$ and integrability}
\label{subsectionproof4}

In the previous subsections we have established higher symmetries of the leading singularities $r_n$ in the dual conformal frame $x_0 \to \infty$. In this and the next subsections we perform a similar analysis directly for $R_n$.

Before doing so, let us comment on a rather surprising connection. It turns out that the formula for the leading singularities $R_n$ resembles very closely a special case of a Grassmannian formula for tree-level Amplituhedron volume functions considered in \cite{Ferro:2015grk,Ferro:2016zmx} (see also \cite{Bai:2015qoa}). Indeed, consider eq. (2.4) of \cite{Ferro:2016zmx} for the volume functions $\Omega_{n,k}^{(m)}(Y,Z)$. Upon setting $k=m=2$ there and re-interpreting their $Y_{\alpha}^{I}$ as $Y_{\alpha=1}^{I} = (Z_A)^{I}$ and $Y_{\alpha=2}^{I} = (Z_B)^{I}$, this formula equals our formula for $R_n$.
Very interestingly, the authors of \cite{Ferro:2015grk,Ferro:2016zmx} show that certain derivatives w.r.t. $Y$ of $\Omega_{n,k}^{(m)}(Y,Z)$ satisfy Yangian differential equations. Moreover, the latter are related to so-called Capelli differential equations \cite{Ferro:2014gca}. 
We comment on how these equations look like in our notations for $R_n$ below.

Momentum twistors are the most appropriate variables to formulate the symmetries of $R_n$, so we write
\begin{align}
R_n = R_n(Z_1,\ldots,Z_n;Z_A,Z_{B}) \,.
\end{align}
We introduce the weight-counting operator 
\begin{align}
H_a := Z_a^J \frac{\pa}{\pa Z^J_a} \,, \label{Hcount}
\end{align}
which commutes with the dual-conformal transformations.
Let us recall that $R_n$ carries nonzero dual-conformal weight with respect to $Z_A$ and $Z_{B}$,
\begin{align}
H_A \, R_n = -4 R_n \; , \quad
H_{B} \, R_n = -4 R_n \; , \quad
H_i \, R_n = 0 \; , \quad i = 1,\ldots, n \,. \label{RnWeight}
\end{align}
Instead of dealing with the $su(2,2)$ generators, it will be more convenient for us to supplement the dual-conformal algebra with the central element $H$, thus we effectively deal with the $gl(4) $ symmetry. Namely, the local generators $\mathfrak{J}_a$ acting on the $a$-th coordinate take the following form in the momentum twistor variables,
\begin{align}
\left(\, \mathfrak{J}_a  \, \right)^{I}{}_{J} = Z_{a}^I \frac{\pa}{\pa Z_{a}^J} \,. \label{Jlocal}
\end{align}
Here $a=A,B,1,2,\ldots,n$. Momentum twistors $Z_A$ and $Z_B$ specifying position of the Lagrangian and $Z_1,\ldots,Z_n$ specifying the light-contour do not appear on equal footing in the Grassmannian integral representation of $R_n$ \p{RnBosGrInt}. Then, it is reasonable
to distinguish them when studying symmetries of $R_n$. Thus, we define two sets of the dual-conformal generators as sums of the local ones \p{Jlocal} with respect to coordinates $1,\ldots,n$ and $A,B$,
\begin{align}
\mathfrak{J}^{I}{}_{J}  := \sum_{i =1 }^{n} \left(\, \mathfrak{J}_i \, \right)^{I}{}_{J}  \;,\quad
\bar{\mathfrak{J}}^{I}{}_{J}  := \left(\, \mathfrak{J}_A \, \right)^{I}{}_{J} + \left(\, \mathfrak{J}_B \, \right)^{I}{}_{J} \,. \label{JJbar}
\end{align}
The statement of the dual-conformal invariance of $R_n$ takes the following form
\begin{align}
\left[ \mathfrak{J}^{I}{}_{J}  + \bar{\mathfrak{J}}^{I}{}_{J} \right]  R_n = -2 \delta^{I}_J \, R_n \,. \label{DCIRn1}
\end{align}

In order to study the higher-order symmetries we define the Yangian generators of $k$-th level in the same way as in eq.~\p{Jfull} with the local $\mathfrak{J}_a$ from \p{Jlocal},
\begin{align}
\left[  \mathfrak{J}^{(k)} \right]^{I}{}_J = \sum_{1 \leq i_1 < i_2 < \ldots < i_{k+1} \leq n} 
\left( \mathfrak{J}_{i_1} \right)^{I}{}_{L_1} 
\left( \mathfrak{J}_{i_2} \right)^{L_1}{}_{L_2} \ldots \left( \mathfrak{J}_{i_{k+1}} \right)^{L_{k}}{}_{J} 
\,. \label{JkExpl}
\end{align}
They act nontrivially on the variables of the Wilson loop contour, but the coordinates $Z_A,Z_B$ of the Lagrangian are inert with respect to them.
Let us recall that the level-zero generators are the dual-conformal generators from eq.~\p{JJbar}, i.e. $\mathfrak{J}^{(0)} = \mathfrak{J}$, and we use the shorthand notations $\widehat{\mathfrak{J}}$ for the level-one Yangian generators,
\begin{align}
\widehat{\mathfrak{J}}^{I}{}_{J} := \left[  \mathfrak{J}^{(1)} \right]^{I}{}_J = \sum_{1\leq i<j\leq n} 
\left( \mathfrak{J}_i \right)^{I}{}_{K} 
\left( \mathfrak{J}_j \right)^{K}{}_{J}  \,. \label{Jhat2}  
\end{align}
We derive higher-order symmetries in Appendix \ref{App:proofsRn}. Let us summarize our findings here.   
We find that the leading singularities $R_n$ are annihilated by the level-two and higher-level Yangian generators
\begin{align}
\left[ \mathfrak{J}^{(k)} \right]^{I}{}_{J} \, R_n = 0 \;,\qquad k=2,\ldots,n-1 \,. \label{JkRn}
\end{align}
The level-one Yangian generators $\widehat{\mathfrak{J}}$ do not annihilate $R_n$. Nevertheless, there are several symmetry relations involving them. Projecting $\widehat{\mathfrak{J}}^{I}{}_{J}$ with the tensor $\vev{Z_A, Z_{B},*,\star}$ we achieve the invariance,
\begin{align}\label{YangianRn}
\ep_{I K L M} Z^K_A Z^L_{B} \,  \widehat{\mathfrak{J}}^{I}{}_{J} \, R_n =  \ep_{J K L M} Z^K_A Z^L_{B} \, R_n \,.
\end{align}
In other words, certain linear combinations of the level-one Yangian generators $\widehat{\mathfrak{J}}$ are symmetries of $R_n$.

Another possibility to arrive at a symmetry relation with the level-one generators is to introduce the dual-conformal variations $\bar{\mathfrak{J}}$ \p{JJbar} which act on the Lagrangian coordinate,
\begin{align}
\left[ \bar{\mathfrak{J}}^{I}{}_{K}\widehat{\mathfrak{J}}^{K}{}_{J} + 2 \widehat{\mathfrak{J}}^{I}{}_{J} + \mathfrak{J}^{I}{}_{J} \right] R_n = 0  \,. \label{JbarJhat}
\end{align}
This relation is equivalent to eq. (4.2) of \cite{Ferro:2016zmx} found for the tree-level Amplituhedron volume function.

Let us also notice that projecting \p{JbarJhat} with the tensor $\vev{Z_A, Z_{B},*,\star}$ we reproduce \p{YangianRn}.
Let us also note that taking the trace of the level-one Yangain generators leads to an invariance, 
\begin{align}
\widehat{\mathfrak{J}}^{I}{}_{I}  \, R_n = 0  \,. \label{JhatTr}
\end{align}
In fact, the latter relation is a direct consequence of the dual-conformal invariance \p{DCIRn1}. 

Besides eq.~\p{YangianRn}, we also find another bi-local combination of the dual-conformal generators which annihilates $R_n$,
\begin{align}
\left[ \left( \mathfrak{J}_i \right)^{K}{}_{J} \, \bar{\mathfrak{J}}^{J}{}_{I} + 4 \left( \mathfrak{J}_i \right)^{K}{}_{I}  \right] R_n = 0 \, , \qquad i = 1,\ldots,n \,. \label{bonusDE}
\end{align}
or more explicitly,
\begin{align}
\left[ \frac{\pa}{\pa Z_i^J} Z_A^J \frac{\pa}{\pa Z_A^I} + \frac{\pa}{\pa Z_i^J} Z_{B}^J \frac{\pa}{\pa Z_{B}^I}  + 4 \frac{\pa}{\pa Z_i^I} \right] \, R_n = 0 \, , \qquad i = 1,\ldots,n \,.
\end{align}
Similar to eq.~\p{bonusDiffEq}, we can trace back the origin of \p{bonusDE} into the cyclic shift transformations of the Yangain. Indeed, the cyclic shift $Z_i \to Z_{i+1}$ does not change the Grassmannian representation for $R_n$, see eq.~\p{RnBosGrInt}. However, it changes the Yangian generators $\widehat{\mathfrak{J}}$ \p{Jhat2}. Consistency of eqs.~\p{YangianRn} and  \p{JbarJhat} under the cyclic shift is garanteed by \p{bonusDE}.

We notice that the found symmetries \p{JkRn}, \p{YangianRn}, \p{JbarJhat}, \p{JhatTr}, \p{bonusDE} of $R_n$ are consistent with those of $r_n$ from Sect.~\ref{sec:symm_rn}.
Choosing $Z_A \wedge Z_{B}$ as the infinity bi-twistor brings $R_n$ down to $r_n$, see eq.~\p{rnRn}, and the symmetry relations for $R_n$ presented in this Section boil down to those for $r_n$ from Sect.~\ref{sec:symm_rn}. This is almost trivial for those symmetry relations which do not involve differentiation of the Lagrangian coordinate, namely those without $\bar{\mathfrak{J}}$. Indeed, the level-$k$ (with $k\geq 2$) Yangian invariances of $R_n$ in eq.~\p{JkRn} turn into those of $r_n$ in eq.~ \p{eq:main-Yangian-rn3}. The trace of level-one generators in eq.~\p{JhatTr} is consistent with values of traces in eq.~\p{eq:main-Yangian-rn2}. The frame $Z_A \wedge Z_{B} \to \ep^{\da\db}$ choice in eq.~\p{JkRn} restricts indices $I = \alpha$ and $J=(\beta,\dot\beta)$ that leads to the first two eqs. in \p{eq:main-Yangian-rn2}.
Eq.~\p{bonusDE} implies eq.~\p{bonusDiffEq}. In order to see it, we eliminate $\bar{\mathfrak{J}}$ in \p{bonusDE} owing to the dual-conformal invariance \p{DCIRn1} of $R_n$, and then we can choose $Z_A \wedge Z_{B}$ as the infinity bi-twistor. Likewise, we can show that 
eq.~\p{JbarJhat} implies the four eqs. in \p{JJhat1} and \p{JJhat2}.

We delegated the proofs of the relations given above to Appendix~\ref{App:proofsRn}. Similar to considerations in subsection \ref{subsectionproof1}, they rely on the quantum-inverse scattering method constructions \cite{Faddeev:1996iy}. There are slight modifications w.r.t. subsection \ref{subsectionproof1} to take into account that $R_n$ carries nonzero dual-conformal weights \p{RnWeight}.

%
%
%

\section{Results for $F_n$ in perturbation theory}
\label{sec:perturbative_results}

In this section we provide evidence that the twistor dual-conformal functions $B$ \p{I5ptBos} and \p{I4ptBos}, or equivalently their space-time versions $b$ \p{Cfunc} and \p{Bfunc}, provide a basis of leading singularities for $F_n$ at higher loops. This is important both in view of future bootstrap applications and a supersymmetric generalization of the observable \cite{CHTinpreparation}.

\subsection{Result for $F_n$ at one loop}

\begin{figure}[t]
\begin{center}
\includegraphics[width=0.55\columnwidth]{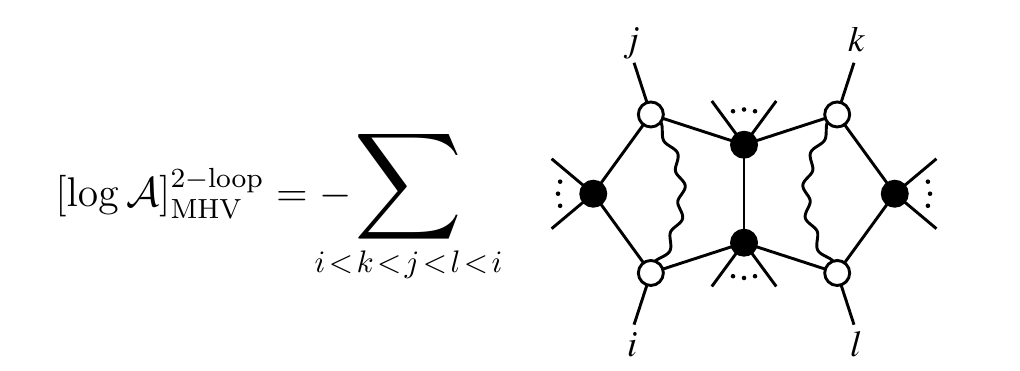}
\end{center}
\caption{Two-loop logarithm of the MHV amplitude. Figure from ref.~\cite{ArkaniHamed:2010gh}.}
\label{twolooplogA}
\end{figure}

\begin{figure}[t]
\begin{center}
\includegraphics[width=0.25\columnwidth]{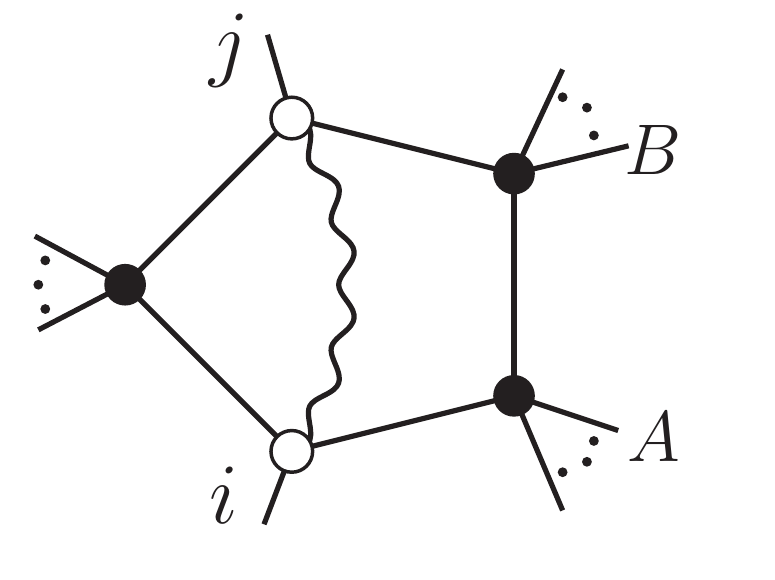} \qquad\qquad
\includegraphics[width=0.2\columnwidth]{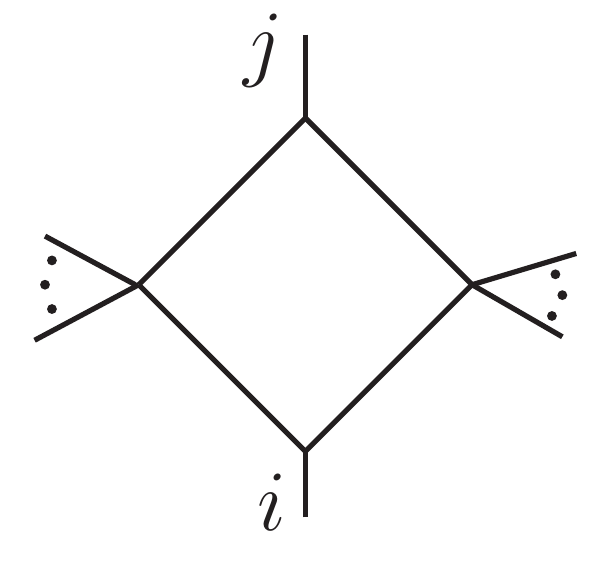}
\end{center}
\caption{The chiral pentagon one-loop integral $I^{\rm cp}_{ij}$ \p{generalpentagon}, on the left. The one-loop box in the two-mass easy configuration $F^{\rm 2me}_{ij}$ \p{F2me}, on the right.}
\label{IcpFig}
\end{figure}

In ref.~\cite{ArkaniHamed:2010gh} a compact expression for the logarithm of MHV amplitudes is given, see Fig.~\ref{twolooplogA}. From this expression it is easy to read off an expression for $F_n$ in a manifestly finite way.
The formula reads\footnote{We explicitly symmetrized the two-loop formula in $AB$ and $CD$.} 
\begin{align}
F^{(1)}_{n} = -\frac{1}{2} 
\sum_{1\leq i < k < j <l \leq n} 
\left[ 
I^{\rm cp}_{ij} LS_{ijkl} + 
 I^{\rm cp}_{kl} LS_{klij}
  \right] \,, \label{F11loop}
\end{align}
where the rational factors $LS$ are the leading singularities,  which depend on a number of twistors specifying the Wilson loop contour and the Lagrangian insertion at $(AB)$,
\begin{align}\label{LSform1}
LS_{ijkl} := \frac{\vev{ijkl} \vev{AB|(k-1 k k+1) \cap (l-1 l l+1)}}{\vev{ijAB}\vev{AB k-1 k} \vev{AB k k+1} \vev{AB l-1 l} \vev{AB l l+1}} \,,
\end{align}
and the dual-conformal chiral one-loop pentagon $I_{\rm cp}$ is a finite pure function, Fig.~\ref{IcpFig},
\begin{align}
I^{\rm cp}_{ij}  = \int_{CD} \frac{ \vev{CD|(i-1 i i+1) \cap (j-1 j j+1)} \vev{ijAB} }{ \vev{CDi-1 i}\vev{CD i i+1} \vev{CD j-1 j} \vev{CDj j+1} \vev{CDAB} }\,. \label{generalpentagon0}
\end{align}
Here, starting from the double pentagon integrand in Fig.~\ref{twolooplogA}, a factor $1=\vev{ijAB}/\vev{ijAB}$ was split between the leading singularity and the pentagon integral, in order to make the latter a pure function.
The chiral pentagon integral evaluates to a pure function
\begin{align}\label{generalpentagon}
I^{\rm cp}_{ij}  =& -\frac{1}{2} \log^2 \left( u_3^{ij} \right) - {\rm Li}_{2} \left(1-u_1^{ij}\right) - {\rm Li}_{2}\left(1-u_2^{ij}\right) \nonumber \\ & - {\rm Li}_{2}\left(1-\frac{u_1^{ij}}{u_3^{ij}}\right)  - {\rm Li}_{2}\left(1-\frac{u_2^{ij}}{u_3^{ij}}\right)  + {\rm Li}_{2}\left(1-\frac{u_1^{ij} u_2^{ij}}{u_3^{ij}}\right)  \,,
\end{align}
of the three cross-ratios
\begin{align}
u_1^{ij} =& \frac{\vev{i-1 i j j+1}\vev{i i+1 AB}}{\vev{i-1 iAB}\vev{i i+1 j j+1}} \,, \notag\\ u_2^{ij} =& \frac{\vev{i i+1 j-1 j}\vev{j j+1 AB}}{\vev{i i+1 j j+1}\vev{j-1 jAB}} \,, \notag\\ u_3^{ij} =& \frac{ \vev{i-1 i j-1 j}\vev{i i+1AB}\vev{j j+1AB}}{\vev{i-1 iAB}\vev{i i+1 j j+1}\vev{j-1 jAB}} \,. \label{u1u2u3}
\end{align}
So we have obtained a closed form analytic expression for $F^{(1)}_n$.

The formula for the pentagon is slightly rewritten from the form given in \cite{ArkaniHamed:2010gh}, so that boundary terms from \p{F11loop} with $u_1$ or $u_2$ vanishing can be considered without using dilogarithm identities. Let us note that the pure function \p{generalpentagon0} is the two-mass-easy one-loop box function $F^{\rm 2me}_{ij}$. Indeed, in the frame $x_0 \to \infty$, or equivalently $Z_A \wedge Z_{B} \to \ep^{\da\db}$, the dual-conformal chiral pentagon turns into the one-loop box in the easy-mass configuration of the external legs, Fig.~\ref{IcpFig},
\begin{align}
F^{\rm 2me}_{ij} := I^{\rm cp}_{ij}\left( u_1^{ij} = \frac{x_{i-1 \, j}^2}{x_{ij}^2}, u_2^{ij} = \frac{x_{i\, j-1}^2}{x_{ij}^2}, u_3^{ij} = \frac{x_{i-1 \, j-1}^2}{x_{ij}^2} \right) . \label{F2me}
\end{align}
The 'magic' numerator of \p{generalpentagon} turns into a pair of chiral fermion propagators, which suppress IR-divergences of the box integral.

\begin{figure}[t]
\begin{center}
\includegraphics[width=0.35\columnwidth]{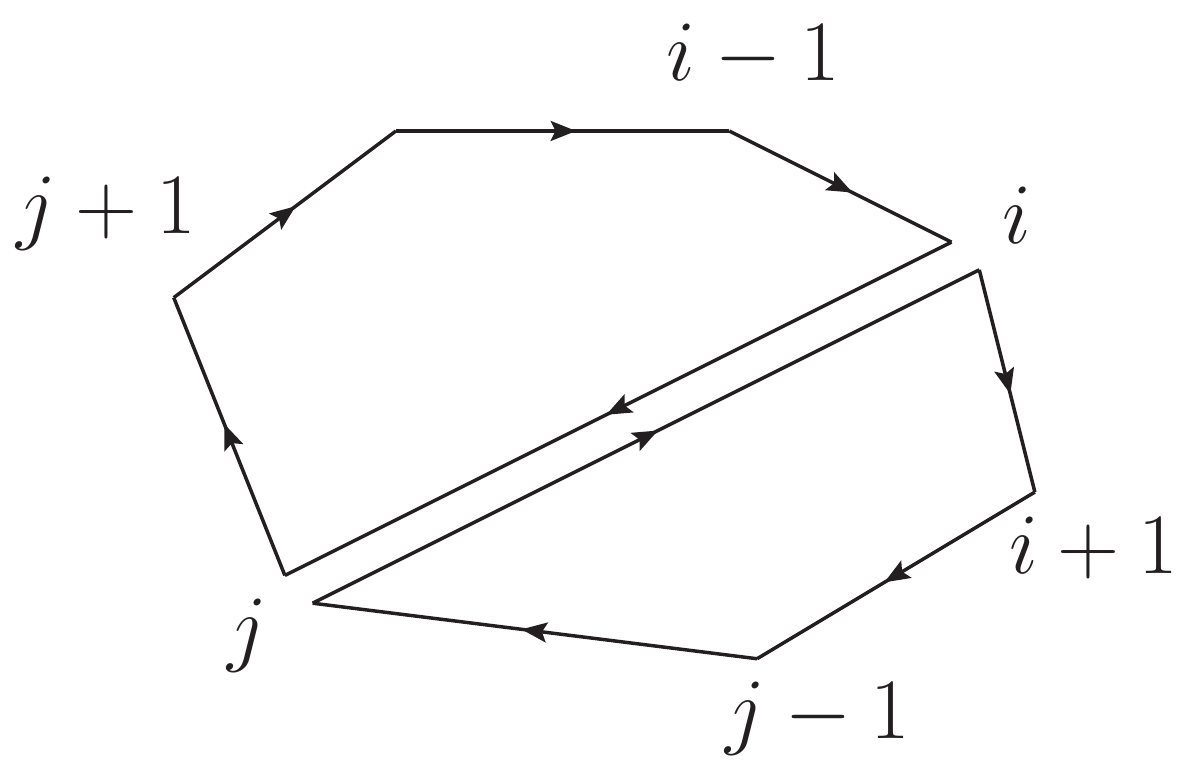}
\end{center}
\caption{The three terms in eq.~\p{F1n} correspond to the three polygons on the figure.}
\label{pic:2polygons}
\end{figure}

We would like to see whether the formula for the one-loop $n$-point $F_n^{(1)}$ can be expressed in terms of the conformal invariants $B$ \p{I5ptBos} and \p{I4ptBos} introduced earlier. 
In order to achieve this, we rewrite coefficients of the pentagon integrals $I_{\rm cp}$ in terms of the $k$-point functions $F_k^{(0)}$ \p{FnBosTree}, which are invariant upon the cyclic shift of their arguments. We find,
\begin{align}
F_n^{(1)} = \sum_{\substack{1\leq i<j \leq n \\ 1 < |i-j|< n-1 }} \biggl[
& F_n^{(0)}(1,2,\ldots,n) - F^{(0)}_{j-i+1}(i,i+1,\ldots,j-1,j) \notag\\ 
& - F^{(0)}_{n+1+i-j}(j,j+1,\ldots,n-1,n,1,\ldots,i-1,i)  \biggr] I^{\rm cp}_{ij} \,. \label{F1n}
\end{align}

In order to explain notations, let us graphically interpret eq.~\p{F1n}. Consider a polygon with vertices $1,2,\ldots,n$ in Fig.~\ref{pic:2polygons}, and attach twistor $Z_k$ to each vertex $k$. Draw the diagonal $(i,j)$. The diagonal splits the polygon into two sub-polygons: $i,i+1,\ldots,j-1,j$ and $j,j+1,\ldots,n-1,n,1,2,\ldots,i-1,i$. The three terms in \p{F1n} correspond to the three polygons. More precisely, we evaluate the twistor function \p{FnBosTree}, which is a linear combination of $B$'s, on the set of cyclically ordered twistors specified by the polygons. Let us stress that since the diagonal $(ij)$ is not light-like, then the last two terms in \p{F1n} could not be interpreted as the Born-level observables; instead, they are understood as $(j-i+1)$-point and $(n+1+i-j)$-point instances of \p{FnBosTree}.
Nevertheless, since \p{FnBosTree} is a linear combination of the four- and five-point $B$-functions, all leading singularities in $F_n^{(1)}$ can be expressed in terms of the $B$ functions. Let us also note that \p{FnBosTree} is invariant under cyclic shifts, and the three terms in the square brackets in \p{F1n} are invariant under cyclic shifts. We explicitly checked eq.~\p{F1n} up to $n=15$.

Let us write out the results \p{F1n} for five and six points explicitly. 
At $n=5$, we find
\begin{align}
F_{5}^{( 1)} =& F_5^{(0)} \, \left( I^{\rm cp}_{13} + I^{\rm cp}_{14} + I^{\rm cp}_{24} + I^{\rm cp}_{25} + I^{\rm cp}_{35} \right)
\notag\\ 
&
- B_{1234} I^{\rm cp}_{14}  - B_{2345} I^{\rm cp}_{25} - B_{1345} I^{\rm cp}_{13} - B_{1245} I^{\rm cp}_{24} - B_{1235} I^{\rm cp}_{35} \,, \label{F5oneloop}
\end{align}
where the Born-level $F_5^{(0)}$ expression is given
in terms of the $B$ functions in eq. \p{F50}.

At $n=6$, we have the following explicit expression for the Born-level $F_6^{(0)}$, see eq.~\p{FnBosTree},
\begin{align}
F_6^{(0)}  = B_{1256} + B_{2356} + B_{3456} - B_{12346} - B_{12456} - B_{23456}  \,.
\end{align} 
The one-loop result can be written as follows,
\begin{align}
F_{6}^{( 1)} = & F_6^{(0)} \left( I^{\rm cp}_{13} + I^{\rm cp}_{14} + I^{\rm cp}_{15} + I^{\rm cp}_{24} + I^{\rm cp}_{25} + I^{\rm cp}_{26} + I^{\rm cp}_{35} + I^{\rm cp}_{36} + I^{\rm cp}_{46}  \right) \notag\\
& - \left( B_{1234} + B_{1456} \right) I^{\rm cp}_{14} - \left( B_{1256} + B_{2345} \right) I^{\rm cp}_{25}  -  \left( B_{1236} + B_{3456} \right) I^{\rm cp}_{36} \notag\\
& + \left( B_{13456} - B_{1356} - B_{3456} \right) I^{\rm cp}_{13} + \left( B_{12345} - B_{1245} - B_{2345} \right) I^{\rm cp}_{15} \notag\\
& + \left( B_{12456} - B_{1256} - B_{2456} \right) I^{\rm cp}_{24}  + \left( B_{23456} - B_{2356} - B_{3456} \right) I^{\rm cp}_{26} \notag\\
&  + \left( B_{12356} - B_{1256} - B_{2356} \right) I^{\rm cp}_{35} + \left( B_{12346} - B_{1246} - B_{2346} \right) I^{\rm cp}_{46} \,.
\end{align}
We hope that these expression may prove useful in future investigations into the structure of the loop corrections for this interesting observable.

\subsection{Structure of the two-loop five-point result}

Although the Grassmannian formula was derived for  $F_{n}^{(0)}$ only, one may hope that this formula also covers the leading singularities at loop level. We have already seen that this is the case for $F_{n}^{(1)}$. Here we mention further evidence for this for $n=5$ at two loops.

In \cite{CHtoappear2} we consider five-cusp planar observable $F_{5}$ in the two-loop approximation, namely $F_{5}^{( L)}$ with $ L =2$ in eq.~\p{FnPertSer}. The explicit perturbative calculations reveals that
\begin{align} \label{F5Lloops}
F_{5}^{( L)} =& F_5^{(0)}\, g^{( L)}_0 ({\bf u})  \\ &+ B_{1234} g^{( L)}_1 ({\bf u}) + B_{2345} g^{( L)}_2 ({\bf u}) + B_{1345} g^{( L)}_3 ({\bf u}) + B_{1245} g^{( L)}_4 ({\bf u}) + B_{1235} g^{( L)}_5({\bf u}) \,,\nonumber
\end{align}
where $g_s^{( L)}({\bf u})$ with $s=0,1,\ldots,5$ are pure functions of transcendental weight $2 L$ of four dual-conformal invariant cross-ratios build from $x_1,\ldots,x_5,x_0$. 
There are four independent cross-ratios. They can be chosen as follows
\begin{align}
{\bf u} = \left\{ \frac{x_{13}^2 x_{20}^2 x_{50}^2}{x_{25}^2 x_{10}^2 x_{30}^2} \;,\;
\frac{x_{14}^2 x_{20}^2 x_{50}^2}{x_{25}^2 x_{10}^2 x_{40}^2} \;,\;
\frac{x_{24}^2 x_{50}^2}{x_{25}^2 x_{40}^2} \;,\;
\frac{x_{35}^2 x_{20}^2}{x_{25}^2 x_{30}^2}  \right\}\,.
\end{align} 
The pure functions $\{ g_{s}^{( L)}({\bf u}) \}_{s=1\ldots 5}$, as well as the accompanying four-point rational prefactors $B_{1234}, B_{2345} , B_{1345}, B_{1245} , B_{1235}$, are related by cyclic shifts: $g_{s+1}^{( L)} = g_{s}^{( L)}|_{Z_i \to Z_{i+1}}$. The Born-level $F_5^{(0)}$, which was given in eq.~\p{F50} in terms of the $B$'s, is cyclically-symmetric. Let us recall that eq.~\p{F5oneloop} is a one-loop instance $L = 1$ of eq.~\p{F5Lloops}.

\section{Connection to all-plus amplitudes in pure Yang-Mills theory}
\label{sect:all-plus-YM}

The light-like polygon geometry is dual to massless scattering kinematics. One might wonder whether there is a closer analogy to physical scattering processes. We find that this is indeed the case. 
As we will show presently, the  leading singularities $r_n$ appear in $n$-gluon scattering in the all-plus helicity configuration in pure Yang-Mills theory. What is more, we find a relationship connecting $F_{n}^{(L)}$ and the $(L+1)$-loop all-plus amplitude in pure Yang-Mills theory.

\subsection{Matching of leading singularities and one-loop all-plus amplitude}

The all-plus amplitude in pure Yang-Mills theory vanishes at tree-level. At one loop, it is a rational function of helicity spinors \cite{Bern:1993qk}. At two loops, it contains both polylogrithmic and rational terms, with the the maximal transcendental weight being two \cite{Badger:2019djh,Dunbar:2016cxp}. 

In \cite{Mafra:2012kh,He:2015wgf,Henn:2019mvc} a concise representation of the one-loop all-plus amplitude was given as the sum of $C_{kmn}$ functions defined in eq.~(2.9) of that ref. \cite{Henn:2019mvc}. These functions where found to be conformally invariant, and moreover have a certain directional dual conformal symmetry.

This motivates us to look for a connection between these coefficients, and the conformal invariants discussed in this paper. To match the kinematics, we first go to the frame $x_0 \to \infty$.
We then find that the functions $C_{kmn}$ with $3\leq k \leq n-1$ and $k+1 \leq m \leq n$ are linear combinations of the conformal invariants $\{ {\rm PT}_n \cdot b_{ijkl} \}$ and $\{ {\rm PT}_n \cdot b_{ijklm} \}$ defined in the present work in eqs.~\p{Cfunc} and \p{Bfunc} which are normalized with the Parke-Taylor factor \p{PT}.

In particular, this means that the functions $C_{kmn}$ have a contour integral representation as discussed in Section~ \ref{subsection:grassmannians},
and all symmetry statements from Section~\ref{subsectionproof1} hold for $C_{kmn}$.

Since the one-loop all-plus $n$-gluon YM amplitude \cite{Bern:1993qk} is a linear combination of conformal invariants $C_{kmn}$, we can rewrite it in terms of the $b$ invariants of the present paper. 
We find the following explicit formula for the one-loop color-ordered partial amplitude $A^{(1)}_{{\rm YM},n}$
\footnote{\label{footnoteNormYM} For the sake of simplicity, we slightly changed normalization of the amplitude as compared to that of Refs.~\cite{Bern:1993qk} and \cite{Dunbar:2016cxp}, 
$ A^{(L) {\rm Ref.}\text{\cite{Bern:1993qk}}}_{{\rm YM},n} 
= - \frac{i}{3} A^{(L) {\rm our}}_{{\rm YM},n}$.},
\begin{align}\label{eq:YMallplus1loopintermsofb}
 A^{(1)}_{\text{YM},n}  = 
 {\rm PT}_n  \, a_n(1,2,\ldots,n) \,, 
\end{align}
where the linear combination $a_n$ of the conformal invariants $b$ is completely analogous to the linear combinations of the dual-conformal $B$'s in \p{FnBosTree},
\begin{align}
 a_n(1,2,\ldots,n) =  \sum_{i=1}^{n-3} b_{i,i+1,n-1,n} - \sum_{i=1}^{n-4} b_{i,i+1,n-2,n-1,n} - \sum_{k=2}^{n-4}\sum_{i=1}^{n-k-3}b_{i,i+1,i+k,i+k+1,n} \,. \label{AllPlusGr}
\end{align}
Eq. (\ref{eq:YMallplus1loopintermsofb}) allows us to make two interesting observations:
Firstly, 
taking into account expression \p{FnBosTree} for $F_n^{(0)}$ in terms of the $B$'s, we find that
the one-loop all-plus $n$-gluon YM amplitude is obtained by taking the limit $x_0 \to 
\infty$ in the tree-level observable $F^{(0)}_n$, 
\begin{align}
A^{(1)}_{\text{YM},n}  = {\rm PT}_n \cdot F_n^{(0)}\biggr|_{Z_A \wedge Z_{B} \to \ep^{\da\db}} \,. \label{YM1loopF}
\end{align}
Secondly, recalling the contour integral representation \p{confGr} of the conformal invariants, we
could say that the all-plus one-loop Yang-Mills amplitude corresponds to 
\begin{align}
A^{(1)}_{\text{YM},n} =  {\rm PT}_n \int_{\Gamma \subset {\rm Gr}(2,n)}
\hspace{-1.0cm}  D^{2(n-2)}C \,
\frac{\left( \sum\limits_{i < j} (ij) \vev{\la_i|x_i \tilde x_j |\la_j} \right)^2 }{(12)(23) \ldots (n1)}  \prod_{r=1,2}
 \delta^{2} \left( \sum_{i=1}^n c_i^r \la_i \right)  , \label{confGrAmpl}
\end{align}
for an appropriate choice of integration cycle $\Gamma$ on the Grassmannian manifold ${\rm Gr}(2,n)$.

\subsection{Connection to two-loop all-plus amplitudes}

Is the resemblance of the all-plus Yang-Mills amplitude and the observable $F_n$  restricted to the lowest perturbative approximation?
To answer this question, let us proceed and compare our analytic one-loop result for $F_n^{(1)}$ \p{F1n} to the two-loop all-plus Yang-Mills amplitude. Ref.~\cite{Dunbar:2016cxp}
contains the polylogarithmic part of the leading-color two-loop all-plus $n$-particle amplitude in Yang-Mills theory. More precisely, the leading-color $ L$-loop partial amplitude $A_{{\rm YM},n}^{( L)}$ evaluated in dimensional regularization $D = 4 - 2 \ep$ is defined in eqs. (1.1) and (1.2) of that paper, and the two-loop amplitude has the following expansion in the dimensional regularization parameter,
\begin{align}
A_{{\rm YM},n}^{(2)} = A_{{\rm YM},n}^{(1)} I_n^{(1)} + P_{{\rm YM},n}^{(2)} + R_{{\rm YM},n}^{(2)} +{\cal O}(\ep) \label{A2loopExpand}
\end{align}
where $I_n^{(1)}$ contains dipole infrared divergences (see also \cite{Catani:1998bh}),
\begin{align}
I_n^{(1)}  = - \frac{1}{\ep^2}\sum_{i=1}^n \left( \frac{\mu^2}{ - 2 p_i \cdot p_{i+1}}\right)^\ep + \frac{n \pi^2}{12} \,. \label{Ipoles}
\end{align}

The finite part consists of the polylogarithmic $P_{{\rm YM},n}^{(2)}$ and rational $R_{{\rm YM},n}^{(2)}$ terms. The transcendental terms $P_{{\rm YM},n}^{(2)}$ are given in eq. (1.9) of Ref.~\cite{Dunbar:2016cxp}. We find that the finite maximally transcendental piece of the two-loop amplitude can be rewritten in the following suggestive form (we changed normalization of the amplitude as given in footnote \ref{footnoteNormYM})
\begin{align}
P_{{\rm YM},n}^{(2)} = & 
- {\rm PT}_n  \sum_{\substack{1\leq i<j \leq n \\ 1 < |i-j|< n-1 }} \biggl[
a_{j-i+1}(i,i+1,\ldots,j-1,j) \notag\\
&  + a_{n+1+i-j}(j,j+1,\ldots,n-1,n,1,\ldots,j-1,j)  \biggr] F^{\rm 2me}_{ij} \label{P2YM} \,,
\end{align}
with the two-mass-easy function $F^{\rm 2me}_{ij}$ \p{F2me}. Let us note that in the four-point case $n=4$ the coefficient in the square bracket is trivially zero and $P_{{\rm YM},4}^{(2)} = 0$ as expected~\cite{Bern:2002tk}.

For the sake of clarity, let us mention that the terms $a_{j-i+1}$ and $a_{ n+1+i-j}$ in eq.~\p{P2YM} are $(j-i+1)$-point and $(n+1+i-j)$-point instances of the function \p{AllPlusGr}, respectively, evaluated along the two subpolygons in Fig.~\ref{pic:2polygons}. The twistor function \p{AllPlusGr} is a linear combination of the $b$'s. Let us stress that $a_{j-i+1}$ and $a_{n+1+i-j}$ from \p{P2YM} cannot be interpreted as the one-loop amplitudes because of the momentum conservation.

We would like to match the polylogarithmic finite part $P_{{\rm YM},n}^{(2)}$ of the YM two-loop amplitude \p{P2YM} with the the one-loop observable $F_n^{(1)}$ \p{F1n}. 

Comparing their accompanying leading singularities in the frame $x_0 \to \infty$, we observe they are almost identical, with the difference being proportional to the one-loop YM amplitude $A^{(1)}_{\text{YM},n}$ \p{YM1loopF},
\begin{align}\label{F2allplusdiscrepancy}
P_{{\rm YM},n}^{(2)} + A^{(1)}_{\text{YM},n} \left(  \sum_{\substack{1\leq i<j \leq n \\ 1 < |i-j|< n-1 }} F^{\rm 2me}_{ij} \right) = {\rm PT}_n \cdot F_n^{(1)}\biggr|_{Z_A \wedge Z_{B} \to \ep^{\da\db}} \,.
\end{align}
This discrepancy can be compensated by a finite IR subtraction, namely by redefinition of $I_n^{(1)}$ \p{Ipoles} in eq.~\p{A2loopExpand}.

Looking more closely at eq. (\ref{F2allplusdiscrepancy}), one realizes that the additional term in parenthesis on the left-hand-side is exactly the (finite part of the) one-loop MHV amplitude. 
Recalling the Wilson loop / scattering amplitude duality, one realizes that this factor appears in the denominator of the definition of $F_n$, cf. eq. (\ref{introdefratio}). In this way we arrive at the following duality relation,
\begin{align}\label{all-plus-duality2}
 {\rm PT}_{n}   \vev{ W_n {\cal L}(x_0)}_{x_0 \to \infty} \sim A^{{\rm YM, all-plus}}_{n} \,.
    \end{align}
This relation is to be understood for the leading transcendental parts of the two objects in the planar limit, up to scheme differences (as usual in Wilson loop / scattering amplitude dualities, these scheme differences arises due to the fact that formally, the Wilson loop has ultraviolet divergences, while the amplitudes have infrared divergences). 
We have just shown that eq. (\ref{all-plus-duality2}) holds up to order $g_{\rm YM}^4$ in perturbation theory, for arbitrary $n$.

Let us present another version of this formula, directly in terms of finite quantities, similar to eq. (\ref{F2allplusdiscrepancy}).
The perturbative expansion of the all-plus Yang-Mills amplitude starts from the one-loop approximation $A^{(1)}_{\rm YM,n}$, which is the rational function given in eq. \p{eq:YMallplus1loopintermsofb}, up to ${\cal O}(\ep)$ corrections. The perturbative expansion of the renormalized planar all-plus amplitude can be written in the following factorized form,
\begin{align}
A_n^{\rm YM} =  {\cal Z}^{\rm YM}_{\rm IR} A^{(1)}_{\rm YM,n} {\cal H}_n^{\rm YM}\,.
\end{align}
Here the finite remainder ${\cal H}_n^{\rm YM}$ is known for $n=4$ up to the three-loop order \cite{Jin:2019nya,Caola:2021izf}. The infrared divergences are taken into account by the factor ${\cal Z}^{\rm YM}_{\rm IR}$, which implements the minimal subtraction. 
Next we invoke the Wilson loop / amplitude duality \p{duality-relation} 
\begin{align}
\vev{W_n} \sim \frac{A_n^{\rm MHV}}{A_{n, \rm tree}^{\rm MHV}} 
=  {\cal Z}^{\rm MHV}_{\rm IR} {\cal H}_n^{\rm MHV}\,,
\end{align}
where the factor ${\cal Z}^{\rm MHV}_{\rm IR}$ minimally subtracts infrared divergences, and ${\cal H}_n^{\rm MHV}$ is the finite remainder. We can identify the pure Yang-Mills and ${\cal N}= 4$ super-Yang-Mills infrared factors, $ {\cal Z}^{\rm MHV}_{\rm IR} \sim {\cal Z}^{\rm YM}_{\rm IR}$. Indeed, in \p{all-plus-duality2}, we compare only the leading transcendental terms of the two objects. The infrared factors coincide in this approximation, in agreement with the maximal transcendentality principle \cite{Kotikov:2002ab}. Taking the logarithm on both sides of eq. \p{all-plus-duality2}, we then obtain the following relation between three finite objects: the finite remainder of the all-plus amplitude, the finite remainder of the MHV amplitude, and the ratio $F_n$, \begin{align}
\log {\cal H}_n^{\rm MHV}+ \log\left(F_n/F_n^{(0)}\right)_{x_0 \to \infty} \sim    \log{\cal H}_n^{\rm YM}  + {\cal O}(\ep) \,.\label{all-plus-duality3}
\end{align}
As before, this is to be understood at the level of the leading transcendental terms. Let us now check eq. (\ref{all-plus-duality3}) for $n=4$.
Note that since $A_{n}^{\rm YM}$ vanishes at tree-level, the quantities on the RHS of eq. (\ref{all-plus-duality3}) enter at one loop order higher than the quantities on its LHS.
Substituting the available perturbative data into this equation, namely the  the ABDK/BDS ansatz \cite{Bern:2005iz,Anastasiou:2003kj} expression for the four-point MHV amplitude, the known expression for $F_4$ up to two loops, and the three-loop all-plus four-point amplitude \cite{Jin:2019nya,Caola:2021izf,Alday:2013ip} into this equation, we find perfect agreement.

Let us discuss these results. Previously, it had been noted that {\it integrands} of MHV amplitudes in ${\cal N}=4$ sYM and the integrands of the all-plus pure YM amplitudes are related. At one-loop, they are related by a dimensional shift \cite{Bern:1996ja}, or, equivalently, the pure YM integrands are obtained from the sYM ones via a particular insertion of 
$(D-4)$-dimensional components of the loop momenta into the integrand.
At two loops, part of the pure YM integrand is again obtained from the sYM one, albeit through a slightly more complicated insertion \cite{Badger:2013gxa,Badger:2016ozq}.

Here, we have provided evidence 
for a connection between three quantities: the $L$-loop observable $F_n^{(L)}$ and the (finite-part of the) $L$-loop MHV amplitude in ${\cal N}=4$ sYM, and the (finite part of the) $(L+1)$-loop all-plus amplitude in pure YM theory. This is a relation between integrated quantities.

\section{Discussion and outlook}
\label{sec:summary}

In this paper, we studied a class of finite observables in maximally supersymmetric Yang-Mills theory. The bosonic observable, which is essentially a Wilson loop with a Lagrangian insertion, had mainly been studied for four-sided Wilson loops. We performed, for the first time, a systematic analysis of the $n$-particle case.

Already at Born level, we uncovered interesting new structures. While the analytic $n$-particle expression has the expected dual conformal symmetry (following from the definition of the correlator in postion space), we found that there is also an unexpected conformal symmetry, i.e. a conformal symmetry in momentum space. 

This conformal symmetry is present at Born level, and also at loop level, where it manifests itself in the leading singularities that multiply the transcendental functions. Conjecturally, the leading singularities are given by a Grassmannian formula, for which we gave two versions, one for general kinamtics, and one in the dual conformal frame $x_0 \to \infty$.

We studied the symmetry properties in the Grassmannian description, which allowed us to prove the conformal symmetry, and to establish several higher-order symmetries related to a Yangian. Curiously, a very similar Grassmannian formula also appeared independently in another context, namely the canonical form for the tree-level Amplituhedron \cite{Ferro:2015grk,Ferro:2016zmx}.

We also computed the full one-loop $n$-particle observable and provided a compact analytic formula for it. This confirms our leading singularity conjecture for this case. The two-loop five-particle case will be discussed in a separate publication \cite{CHtoappear2}.

There are striking similarities between $F_n$ in the dual conformal frame $x_0 \to \infty$ and all-plus amplitudes in pure Yang-Mills theory. We found that for all cases investigated, the leading weight terms of the two objects are dual to each other. The origin of this duality is unknown. If true in general, it could be used to predict the leading transcendental terms of planar all-plus amplitudes in pure Yang-Mills theory: the four-point four-loop case could be deduced from \cite{Mistlberger:2018etf}, and the five-point three-loop case from the results of \cite{CHtoappear2}.
\\

There are a number of open questions and natural extensions of this work:

 1. The hidden conformal symmetry we found in the frame $x_{0}\to \infty$ ought to have a fundamental explanation, which is yet to be found. Moreover, in the case of scattering amplitudes, writing both conformal and dual conformal symmetry generators in the same variables allowed one to see that their closure gives a Yangian algebra \cite{Drummond:2009fd}. What is the full symmetry of the
 Wilson loop with Lagrangian insertion?
 
 2. We find it very intriguing that similar Grassmannian formulas appeared in the context of tree-level Amplituhedron volume functions \cite{Ferro:2015grk,Ferro:2016zmx,Bai:2015qoa}. It would be interesting to find a momentum-space version that would make it possible to study higher Yangian charges directly in momentum space, as discussed in section \ref{App:subsection-momentum-symmetries}.

3. In Ref.~ \cite{Arkani-Hamed:2021iya}, a novel decomposition of $F_4$ in terms of certain `negative geometries' was given, generalizing ideas from the Amplituhedron. In this way, the contributions at different loop orders could be expressed in terms of manifestly finite Feynman integrals. In this way, certain all-order contributions could to be summed and compared to strong coupling results. It would be very interesting to generalize these results to $n>4$ \cite{ACHTinpreparation}.

4. In the case of scattering amplitudes, (broken) superconformal symmetry, in particular the $\overline{Q}$-equation, played an important role in understanding the structure of the perturbative results \cite{Caron-Huot:2011dec}. It would be very interesting to define supersymmetrised version of our observable, and study its $\overline{Q}$ symmetry \cite{CHTinpreparation}.

We are looking forward to seeing progress on these exciting questions in the near future.

\section*{Acknowledgments}

The authors are indebted to Jaroslav Trnka and Nima Arkani-Hamed for multiple discussions and insightful comments, and to Livia Ferro and Simone Zoia for helpful comments on the draft. This research received funding from the European Research Council (ERC) under the European Union’s Horizon 2020 research and innovation programme (grant agreement No 725110), {\it Novel structures in scattering amplitudes}. DC is supported by the French National Research Agency in the framework of the {\em Investissements d’avenir} program (ANR-15-IDEX-02).

\appendix

\section{Leading singularities $B$ in terms of space-time variables}
\label{sec:Appendix:B-spacetime} 

In this Appendix we show how the dual-conformal invariants $B$ from the main text (eqs.~\p{I5ptBos} and \p{I4ptBos}) look like when expressed in terms of space-time variables.

Let us consider as an example $B_{1234}$, and set $n=5$. Firstly, we make the denominator parity-even
\begin{align}
B_{1234}|_{n=5} = \frac{\vev{1234}}{\vev{AB12}\vev{AB23}\vev{AB34}} \cdot \frac{\vev{1234}\vev{AB (\overline{14})}}{\vev{AB14}\vev{AB (\overline{14})}} \label{B1234}
\end{align}
where $(\overline{14}) = (512) \cap (345)$ is the conjugate of $(14)$, see e.g. \cite{ArkaniHamed:2010gh} and eq.~\p{twPlaneInters}.

The twistor four-brackets of the type $\vev{i i+1 j j+1}$ and $\vev{A B i i+1}$ are provided in \p{twbrTox2}. The the parity-even combination in the denominator of \p{B1234} can be written in $x$ coordinates with the help of
\begin{align}
\vev{AB i j}\vev{AB (\overline{i j})} =  & \vev{AB} \vev{i-1 i} \vev{i i+1} \vev{j-1 j}\vev{j j+1} \times \\  & \hspace{0cm}\bigl[ -x_{0 i-1}^2 x_{0 j-1}^2 x_{ij}^2 
 + x_{0 i-1}^2 x_{0 j}^2 x_{ij-1}^2+
x_{0 i}^2 x_{0 j-1}^2 x_{i-1j}^2
-x_{0 i}^2 x_{0 j}^2 x_{i-1
\,j-1}^2 \bigr]\notag \,.
\end{align}

The numerator of \p{B1234} contains both the parity even and the parity odd pieces. They are revealed as follows
\begin{align}
& \vev{AB (\overline{14})} \vev{1234} - \vev{AB14}\vev{2351}\vev{2345} = \vev{AB} ({\rm PT}_5)^{-1} \epsilon_{123450} \,, \label{epsTw} \\
& \vev{AB (\overline{14})} \vev{1234} + \vev{AB14}\vev{2351}\vev{2345} \notag\\ 
&= \vev{AB} ({\rm PT}_5)^{-1} \left( x_{13}^2 x_{24}^2 x_{50}^2 + x_{13}^2 x_{25}^2 x_{40}^2 - x_{14}^2 x_{25}^2 x_{30}^2 + x_{14}^2 x_{20}^2 x_{35}^2 - x_{10}^2 x_{24}^2 x_{35}^2 \right) ,
\end{align}
where ${\rm PT}_5$ is the Parke-Taylor factor \p{PT}, which balances the spinor helicity weight on both sides of the previous equation.

Substituting the previous identities into eq.~\p{B1234} we obtain space-time expression for one of the dual-conformal invariants at $n=5$, where the numerator is split in the parity-even and parity-odd pieces,
\begin{align}
B_{1234}|_{n=5} = \frac{1}{2}\frac{x_{13}^2(x_{13}^2 x_{24}^2 x_{50}^2 + x_{13}^2 x_{25}^2 x_{40}^2 - x_{14}^2 x_{25}^2 x_{30}^2 + x_{14}^2 x_{20}^2 x_{35}^2 - x_{10}^2 x_{24}^2 x_{35}^2 + \ep_{123450})}{x_{10}^2 x_{20}^2 x_{30}^2(x_{10}^2 x_{35}^2 x_{40}^2 - x_{13}^2 x_{40}^2 x_{50}^2 + x_{14}^2 x_{30}^2 x_{50}^2)}\,.
\end{align}

\section{Details of symmetry proofs for $r_n$}

\subsection{Dual-conformal algebra}
\label{AllDualConf}

The dual-conformal algebra is an instance of $su(2,2)$. The algebra comprises Lorentz rotations split into a pair of symmetric tensors $M_{\a\b}$ and $\bar M_{\da\db}$, dilatation $D$ which counts dimension of the remaining generators, space-time shifts $P_{\a\da}$ and the conformal boost $K_{\a\da}$, which satisfy the following commutation relation
\begin{align}
[K_{\a\da},P^{\db\b}] = \delta^{\beta}_{\alpha} \delta^{\dot\beta}_{\dot\alpha} D + M_{\a}^{\b} \delta^{\dot\beta}_{\dot\alpha}  + \bar M_{\da}^{\db}  \delta^{\beta}_{\alpha} \,.
\end{align}
Representation of the dual-conformal algebra generators in the momentum twistor variables $Z^I=(\la^{\a},\mu^{\da})$ can be read off from the matrix \p{dualConfJmat}. It is also instructive to provide generators in the coordinate space  
\begin{align}
& P_{\a\da} = \frac{\pa}{\pa x^{\da\a}} \;,\quad
K^{\da\alpha} =  x^{\dot\alpha \beta}  x^{\dot\beta\alpha} \frac{\pa}{\pa  x^{\dot\beta\beta}} + x^{\da\beta} \la_{\beta} \frac{\pa}{\pa \la_{\alpha}} \,,\label{dualconfAlgxlaGen0}\\
& D = x_{\a\da} \frac{\pa}{\pa x_{\a\da}} + \frac{1}{2} \la_{\a} \frac{\pa}{\pa \la_{\a}}\;,\quad M_{\a\b} = x_{(\a}^{\dg} \frac{\pa}{\pa x^{\dg \beta)}}+ \la_{(\a} \frac{\pa}{\pa \la^{\beta)}} \;, \quad
\bar M_{\da\db} = x_{(\da}^{\g} \frac{\pa}{\pa x^{\db)\g}}\,. \label{dualconfAlgxlaGen}
\end{align}

\subsection{Variable change in the conformal boost generator} \label{AppKvarChange}

There is a variable change that maps the conformal boost generator $\mathbb{K}_{\a\da}$ \p{KboostDef} to the level-one generator $\widehat{\mathfrak{J}}$ of the Yangian 
\p{JJgenSum} built from the dual-conformal algebra. We used this fact in \p{Kinv} in order to establish the conformal symmetry of $r_n$.
Let us give some more details on this variable change $(\la,\tilde\la) \to (\la, \mu)$, which is given by the following equations,
\begin{align}
& \mu_j^{\da} = x_n^{\da\a} \la_{j\,\a} - \sum_{k=j+1}^n \vev{\la_k \la_j} \tilde\la_k^{\da} \;, \qquad j = 1,\ldots,n-1 \,, \notag \\
& \mu_n^{\da} =  x_n^{\da\a} \la_{n\,\a} \,,
\end{align}
where $x_n$ is fixed.
Differentiations in the helicity spinors $(\la,\tilde\la)$ are reexpressed into derivatives with respect to the twistor variables $Z=(\la,\mu)$ as follows
\begin{align}
& \frac{\pa}{\pa \la_{j\,\a}} = \frac{\pa}{\pa \la_{j\,\a}} -  x_j^{\da\a} \frac{\pa}{\pa \mu_j^{\da}} - \tilde\la_j^{\da}\sum_{k=1}^{j-1} \la_k^{\a} \frac{\pa}{\pa \mu_k^{\da}}\, , \label{der1}\\
& \frac{\pa}{\pa \tilde\la_j^{\da}} 
= - \sum_{k=1}^{j-1} \vev{\la_j \la_k} \frac{\pa}{\pa \mu_k^{\da}}\; , \quad j = 1,\ldots, n \,. \label{der2}
\end{align} 
Substituting \p{der1} and \p{der2} in the second order differential operator of the conformal boost $\mathbb{K}$ \p{KboostDef} one finds how it acts on a function of momentum twistors $f(\la_1,\mu_1,\ldots,\la_n,\mu_n)$ normalized by the Parke-Taylor factor \p{PT}.
The calculation is done in detail in  \cite{Drummond:2010qh}. One finds,
\begin{align}
& \left(\sum_{j=1}^n \frac{\pa^2}{\pa \la_j^{\a}\pa \tilde \la_{j}^{\da}}  \right) \frac{f(\la,\mu)}{\vev{12}\ldots\vev{n1}} \notag \\ 
& = - \frac{1}{\vev{12} \ldots \vev{n1}}\sum_{1\leq j<k \leq n} \left(  \la_{j\,\a} \frac{\pa}{\pa \la_j^\b} \la_k^\b \frac{\pa}{\pa \mu_k^{\da}} + \la_{j\,\a} \frac{\pa}{\pa \mu_j^\db} \mu_k^\db \frac{\pa}{\pa \mu_k^{\da}}  \right) f(\la,\mu) \,, \label{KvarChange}
\end{align}
provided that $f$ is annihilated by the dual-conformal shift $P_{\a\da}$,
\begin{align}
\left(\sum_{j=1}^n \la_i^{\a} \frac{\pa}{\pa \mu^{\da}_j} \right) f(\la,\mu) = 0\,,
\end{align}
and that $f$ is a homogeneous function of the momentum twistor variables with zero weight with respect to each twistor,
\begin{align}
\left( \la_j^{\a} \frac{\pa}{\pa \la_j^{\a}} + \mu_j^{\da} \frac{\pa}{\pa \mu_j^{\da}} \right) f(\la,\mu) = 0 \,, \quad j = 1 , \ldots, n \,.
\end{align}
The leading singularities $r_n$ satisfy both these requirements, thus \p{KvarChange} justifies eq.~\p{Kinv}.

\subsection{Implications of Yangian cyclicity in space-time coordinates}
\label{sec:cyclYangK}

In this Section we provide space-time form of eq.~\p{bonusDiffEq}. As we showed in Sect.~\ref{sec:symm_rn}, this equation originates from the cyclic shift transformations of the Yangian generators. 

Eq.~\p{bonusDiffEq} at $I = \da$ is an immediate consequence of the Poincar\'{e} invariance \p{Poinc}. At $I = \alpha$, eq.~\p{bonusDiffEq} takes the following form,
\begin{align}
\frac{\pa}{\pa \mu_i^{\dot\beta}} \, K^{\dot\beta}{}_{\alpha} \, r_n (Z_1,\ldots,Z_n) = 4 \frac{\pa}{\pa \la_i^{\alpha}} r_n (Z_1,\ldots,Z_n) \;, \qquad i=1,\ldots, n \,. \label{bonusDiffEq2}
\end{align}
with the dual-conformal boost generator in in momentum twistors variables 
\begin{align}
K^{\da}{}_{\a} = \sum_{i=1}^n \mu_{i}^{\da} \frac{\pa}{\pa \la_{i}^{\a}} \,.    
\end{align} 
It is instructive to rewrite relation \p{bonusDiffEq2} in the space-time coordinates,
\begin{align}
\left[ \la_{i+1}^\gamma \frac{\pa}{\pa x_i^{\dot\gamma\gamma}} K^{\dot\gamma\alpha} - 4 \la_{i+1}^{\gamma} x_i^{\dot\gamma \alpha} \frac{\pa}{\pa \tilde x_i^{\dot\gamma\gamma}} + 4 \vev{i \, i+1} \frac{\pa}{\pa \la_{i\,\alpha}} \right] r_n(\la_1,x_1,\ldots,\la_n,x_n) = 0 \; , \label{bonusDiffEqxsapce}
\end{align}
where $K^{\da\a}$ is the dual-conformal boost generator in the space-time coordinate representation, see \p{dualconfAlgxlaGen0},
\begin{align}
K^{\da\alpha} &= \sum_{i=1}^n \left[  x_i^{\dot\alpha \beta}  x_{i}^{\dot\beta\alpha} \frac{\pa}{\pa x_i^{\dot\beta\beta}} +  x_i^{\da\beta} \la_{i \,\beta} \frac{\pa}{\pa \la_{i\,\alpha}} \right] . \label{KBoostSum}
\end{align}
Contracting \p{bonusDiffEqxsapce} with $\la_{i\,\a}$ and taking into account that $r_n$ carries zero helicity weight with respect to each point and the twistor incidence relation $\la_i^{\a}\frac{\pa r_n}{\pa x_i^{\da\a}} = 0 $, we obtain an interesting implication of \p{bonusDiffEq2},
\begin{align}
\frac{\pa}{\pa x_i^{\dot\alpha\alpha}} \left( K^{\dot\alpha\alpha} r_n \right) = 4 x_i^{\dot\alpha\alpha} \frac{\pa}{\pa x_i^{\dot\alpha\alpha}} r_n \;,\quad i=1,\ldots,n\,. \label{bonusDiffEq3}
\end{align}

We can think of  \p{bonusDiffEq3} as an anomaly equation for the dual-conformal symmetry, which is broken by the frame fixing $x_0 \to \infty$, see \p{noK}.

\subsection{From R-operators to the Grassmannian}
\label{subsection:corollaryRoperatorsGrassmannian}

In the Yangian symmetry considerations of the leading singularities $r_n$, we rely upon their $\mathrm{R}$-operator decomposition  \p{goodRseq}. In order to prove this relation, we would like to show that it is equivalent to the Grassmannian representation \p{confGr}. In other words,
\begin{align}
\int  \omega(C) \Delta(C,Z) =  \overrightarrow{\prod_{i=2}^{n-1}} \mathrm{R}_{i+1 i}  \, \cdot\,   \overleftarrow{\prod_{i=1}^{n-2}} \mathrm{R}_{i i+1} \, \ket{\Omega_n} \label{GrR}
\end{align} 
where the arrows specify ordering of the $\mathrm{R}$-operator product; $C$ is a $2\times n$ matrix representing a 2-plane in $\mathbb{C}^n$,
\begin{align}
C = \begin{pmatrix}
c_{1}^1 & \dots & c_n^1 \\
c_{1}^2 & \dots & c_n^2
\end{pmatrix}  := \begin{pmatrix}
\vec c_1 & \dots & \vec c_n  
\end{pmatrix} \label{CMAT}
\end{align}
with $2\times 2$ minors denoted as $(ab)$;
$\omega(C)$ is a measure on the Grassmannian ${\rm Gr}(2,n)$ (see \p{DCmeasure0} and \p{DCmeasure}) for the fixed ordering,
\begin{align}
\omega(C):= \frac{D^{2(n-2)} C}{(12)(23) \ldots (n1)} \,,  \label{omegaC}
\end{align}
and distribution $\Delta(C,Z)$ is defined as
\begin{align}
\Delta(C,Z):= \left( \sum_{i < j} (ij) [\mu_i \mu_j] \right)^2  \prod_{r=1,2}\delta^{2} \left( \sum_{i=1}^n c_i^r \la_i \right) \, . \label{Delta}
\end{align}

We note that $\mathrm{R}_{ij}$ (see eq.~\p{Rop}) acts on $Z_i^{I}$ as the shift $Z_i \to Z_i + t Z_j$. In view of the identity,
\begin{align}
\vec c_i \, (Z_i + t Z_j) + \sum_{k\neq i} \vec c_k \, Z_k =  (\vec c_j + t \vec c_i) Z_j +\sum_{k\neq j} \vec c_k \, Z_k \,, 
\end{align}
the action of $\mathrm{R}_{ij}$ onto $\Delta(C,Z)$ \p{Delta} by the shift of twistor variables is equivalent to the column transformation $\vec c_j \to \vec c_j + t \vec c_i$ of matrix $C$ \p{CMAT}.

Each $\mathrm{R}$-operator in the product \p{GrR} contains one-fold integration (see \p{Rop}). Let us label the corresponding integration variables as $t_1,t_2,\ldots,t_{2n-4}$ from right to the left. Let us start with matrix $C_0$ 
\begin{align}
C_0 = \begin{pmatrix}
1 & 0 & \dots & 0 & 0 \\
0 & 0 & \dots & 0 & 1
\end{pmatrix}  \, \label{C0mat}
\end{align}
which represents a dimension zero cell of the Grassmannian ${\rm Gr}(2,n)$ and corresponds to the pseudo-vacuum state of the spin-chain \p{vacuum},
\begin{align}
\Delta(C_0,Z) = \ket{\Omega_n} \,.
\end{align}
Each $\mathrm{R}$-operator from \p{GrR} implements a column transformation of $C_0$ introducing one extra $t$-parameter and increments dimension of the Grassmannian cell. Acting with all $(2n-4)$ $\mathrm{R}$-operators from \p{GrR} we arrive at the top cell of the Grassmannian. The measure $\omega(C)$ \p{omegaC} takes the canonical form in the $t$-parametrization, and each $\mathrm{R}$-operator \p{Rop} contributes one $d\log$ factor,
\begin{align}
\omega(C) = \frac{dt_1}{t_1} \wedge \frac{dt_2}{t_2} \wedge \ldots  \wedge \frac{dt_{2n-4}}{t_{2n-4}}\,. \label{canon}
\end{align}
The described procedure is equivalent to the BCFW-decomposition and positroid stratification of the Grassmannians in the context of $\cN = 4$ sYM amplitudes \cite{Arkani-Hamed:2016byb}.  

Let us also note that the decomposition \p{GrR} is not unique. We could represent the Grassmanian top cell by another sequence of $\mathrm{R}$-operators. However, not every decomposition is compatible with the monodromy \p{TmatMonodr}. In other words, the Yangian symmetry could be not immediately obvious if we use a decomposition other than \p{GrR}.

It is also instructive to consider the decomposition \p{GrR} at $n=4$,
\begin{align}
r_4 = \underbrace{{\rm R}_{32}}_{t_4} \underbrace{{\rm R}_{43}}_{t_3} \underbrace{{\rm R}_{23}}_{t_2} \underbrace{{\rm R}_{12}}_{t_1} \, \ket{\Omega_4} = 
\int \omega(C_4)\,\Delta(C_4) \label{r4exampl}
\end{align}
where we explicitly indicate the labeling of the one-fold integrations \p{Rop}, and $\ket{\Omega_4} = [\mu_1 \mu_4]^2 \, \delta^2(\la_1) \delta^2(\la_4)$ \p{vacuum}. The sequence of $\mathrm{R}$-operators from \p{r4exampl} corresponds to the following sequence of $C_0$ column transformations \p{C0mat}, such that $\mathrm{R}_{ij}$ acts as $\vec{c}_j \to \vec{c}_j + t \vec{c}_i$,
\begin{align}
& C_0 := \begin{pmatrix}
1 &0 & 0 &0 \\
0 &0 & 0 &1
\end{pmatrix} \overset{{\rm R}_{12}}{\to}
C_1 :=\begin{pmatrix}
1 &t_1 & 0 &0 \\
0 &0 & 0 &1
\end{pmatrix} \overset{{\rm R}_{23}}{\to}
C_2 := \begin{pmatrix}
1 &t_1 & t_1 t_2 &0 \\
0 &0 & 0 &1
\end{pmatrix} \to \\[0.2cm] \notag
& \overset{{\rm R}_{43}}{\to}
C_3 := \begin{pmatrix}
1 &t_1 & t_1 t_2 &0 \\
0 &0 & t_3 &1
\end{pmatrix} \overset{{\rm R}_{32}}{\to}
C_4 := \begin{pmatrix}
1 &t_1(1+t_2 t_4) & t_1 t_2 &0 \\
0 &t_3 t_4 & t_3 &1
\end{pmatrix} \,,
\end{align}
and $C_i$ represents $i$-th dimensional cell of ${\rm Gr}(2,4)$; and $\omega(C_4)$ is the canonical form \p{canon}.

\section{Details of integrability proofs for $R_n$}
\label{App:proofsRn}

In this Appendix we provide proofs of the symmetry relations for the leading singularities $R_n$ presented in Sect.~\ref{subsectionproof4}. The line of argumentation is similar to that of Sect.~\ref{sec:symm_rn} where the Yangian symmetries of $r_n$ are considered, but requires several extra technical details. We split the proof in several steps. In \ref{sec:RLT} we provide the spin-chain formulation of the Yangian algebra generators and define the R-operators. The latter are employed in \ref{sec:RGrRn} to represent $R_n$ as a product of R-operators acting on the pseudo-vacuum state of the spin-chain. In \ref{AppCycl} we discuss the cyclic shift transformations of the Yangian. In \ref{sec:lemmas} we provide several technical statements which are used in the proof of the Yangian invariance relations in \ref{AppRn}.

\subsection{R-operators, Lax-operators, and the monodromy matrix}

\label{sec:RLT}

Let us denote ${\cal H}_i^{(\delta_i)}$ the space of homogeneous functions of $Z_i$ of the homogeneity degree $\delta_i$, namely $H_i = \delta_i$ is the dual-conformal weight \p{Hcount}. An elementary block in the decomposition of the leading singularities $R_n$ will be the R-operator. We will need to introduce the spectral parameter $u$ in the R-operator,
\begin{align}
\left[{\rm R}_{ij}^{(u)} \, G \right](Z_i,Z_j) := \int \frac{dt}{t^{1+u}} G(Z_i + t Z_j,Z_j) \,. \label{Rop2}
\end{align}
The ${\rm R}$-operator \p{Rop2}  acts in the tensor product ${\cal H}_i^{(\delta_i)} \otimes {\cal H}_j^{(\delta_j)}$ and maps it into ${\cal H}_i^{(\delta_i-u)} \otimes {\cal H}_j^{(\delta_j+u)}$. In other words, the ${\rm R}$-operator does not preserve the dual-conformal weight,
\begin{align}
\left[ H_1, {\rm R}_{12}^{(u)} \right] = - u {\rm R}_{12}^{(u)} \,,\qquad
\left[ H_2 , {\rm R}_{12}^{(u)} \right] = u {\rm R}_{12}^{(u)} \,. \label{H12R12}
\end{align}

In order to construct the Yangian generators \p{JkExpl}, we combine the dual-conformal generators \p{Jlocal} in the $4\times 4$ matrix depending on the spectral parameter $u$, and call it the Lax operator,
\begin{align}
\left[ {\rm L}(u) \right]^{I}{}_J = u \,\delta^I_J + Z^I \frac{\pa}{\pa Z^J} \label{Lax2} \,.
\end{align}
The Lax operator is a local object acting in one site of the spin chain. We also decorate the Lax operator with the dual-conformal weight, ${\rm L}_i^{(\delta)}(u)$, to indicate that it acts onto ${\cal H}_i^{(\delta)} \otimes \mathbb{C}^4$. The matrix product of a pair of Lax operators acting in the $i$-th and $j$-th sites has nice exchange relations with the R-operators \p{Rop2},
\begin{align}
& {\rm R}^{(a)}_{ij} \, {\rm L}_i^{(\delta_i)}(u)\, {\rm L}_j^{(\delta_j)}(v)\, = {\rm L}_i^{(\delta_i-a)}(v)\, {\rm L}_j^{(\delta_j+a)}(u)\, {\rm R}^{(a)}_{ij} \,, \qquad a = v-u \label{RLL1}
\end{align} 
and
\begin{align}
& {\rm R}^{(a)}_{ji} \, {\rm L}_i^{(\delta_i)}(u)\, {\rm L}_j^{(\delta_j)}(v)\, = {\rm L}_i^{(\delta_i+a)}(u)\, {\rm L}_j^{(\delta_j-a)}(v)\,{\rm R}^{(a)}_{ji} \,, \qquad a = v-u-\delta_i + \delta_j  \,. \label{RLL2}
\end{align} 
These local exchange relations are exactly what we need to establish the symmetries of $R_n$.

A simple corollary of the exchange relation \p{RLL1} is the dual-conformal invariance \p{Jlocal} of the R-operator 
\begin{align}
\left[ {\rm R}^{(a)}_{ij} , \mathfrak{J}_i + \mathfrak{J}_j \right] = 0  \,. \label{DCIRn2}
\end{align}

We form the monodromy matrix ${\rm T}(u)$ as the matrix product of the Lax operators \p{Lax2} acting on spin-chain sites $1,2,\ldots,n$,
\begin{align}
\left[ {\rm T}(u) \right]^I{}_J:= \left[{\rm L}_1(u)\right]^{I}{}_{K_1} \left[{\rm L}_2(u) \right]^{K_1}{}_{K_2} \ldots \left[{\rm L}_n(u)\right]^{K_{n-1}}{}_J \,. \label{Tmonod2}
\end{align}
The monodromy matrix is a polynomial in the spectral parameter $u$. The coefficients of the polynomial are the Yangian generators \p{JkExpl} acting in the spin-chain sites $1,2,\ldots,n$,  
\begin{align}
\left[ {\rm T}(u)\right]^{I}{}_J = u^n \delta^I_J + u^{n-1} \left[ \mathfrak{J}^{(0)}\right]^{I}{}_J + u^{n-2} \left[\mathfrak{J}^{(1)}\right]^{I}{}_J  + \sum_{k=2}^{n-1} u^{n-k-1} \left[  \mathfrak{J}^{(k)} \right]^{I}{}_J \,, \label{Texpand2}
\end{align}
where we recall $\mathfrak{J}^{(0)} = \mathfrak{J}$ \p{JJbar} and  $\mathfrak{J}^{(1)} = \widehat{\mathfrak{J}}$ \p{Jhat2}.

\subsection{From R-operators to the Grassmannian}
\label{sec:RGrRn}

We will need an analogue of the R-operator decomposition \p{goodRseq} for $R_n$. The presentation in this Section parallels the one in Sect.~\ref{subsection:corollaryRoperatorsGrassmannian}.
We define the pseudo-vacuum of the spin chain which belongs to ${\cal H}^{(-4)}_1 \otimes {\cal H}^{(-4)}_n$,
\begin{align}
\ket{\Omega_n} := \delta^{4}(Z_1)\, \delta^4(Z_n) \,. \label{vac2}
\end{align}
In order to represent the leading singularities $R_n$, we act onto the pseudo-vacuum by the product of $2n$ R-operators \p{Rop2} as follows,
\begin{align}
& R_n = {\rm R}_{AB}^{(-1)} {\rm R}_{BA}^{(-1)} \, \ket{\Phi_n} \,, \label{Rn=RRPhi} \\
& \ket{\Phi_n} := {\rm R}_{1A}^{(-4)} {\rm R}_{nB}^{(-4)} \, \overrightarrow{\prod_{i=2}^{n-1}} \mathrm{R}_{i+1 i}^{(0)}  \, \cdot\,   \overleftarrow{\prod_{i=1}^{n-2}} \mathrm{R}_{i i+1}^{(0)} \, \ket{\Omega_n} \,, \label{Phin}
\end{align}
where for the sake of further convenience we split the product into two factors containing two and $2n-2$ R-operators, and we introduced a shorthand notation $\ket{\Phi_n}$ for the latter product acting on the pseudo-vacuum. The dual-conformal weight counting of the R-operators \p{H12R12} enables us to find the weights of $R_n$ and $\ket{\Phi_n}$,
\begin{align}
R_n,\, \ket{\Phi_n}  \in {\cal H}^{(-4)}_A \otimes {\cal H}^{(-4)}_B \otimes {\cal H}^{(0)}_1 \otimes \ldots \otimes  {\cal H}^{(0)}_n \,. \label{RPhiweights}
\end{align}

The resulting expression in \p{Rn=RRPhi} coincides with the contour integral \p{RnBosGrInt} representation of $R_n$,
\begin{align}
 R_n = \int  \omega(C) \Delta(C,Z)  \,.  
\end{align}
Here we use notations $\omega(C)$ for the measure (see \p{DCmeasure0} and \p{DCmeasure})
\begin{align}
\omega(C):= \frac{(AB)^2\, D^{2n} C}{(12)(23) \ldots (n1)} \,,  \label{omegaC2}
\end{align}
and for the delta-function constraints $\Delta(C,Z)$,
\begin{align}
\Delta(C,Z):= \prod_{r=1,2}\delta^{4} \left( \sum_{i \in \{ A,B,1,\ldots,n \} } c_i^r Z_i \right) \, . \label{}
\end{align}
The pseudo-vacuum state \p{vac2} $\ket{\Omega_n} = \Delta(C_0,Z)$
corresponds to the following $2\times (2n+2)$-matrix, 
\begin{align}
C_0 = \begin{pmatrix}
0 & 0 & 1 & 0 & \dots & 0 & 0 \\
0 & 0 & 0 & 0 & \dots & 0 & 1
\end{pmatrix}  \,. \label{}
\end{align}
Each $\mathrm{R}$-operator in the products \p{Rn=RRPhi} and \p{Phin} contains one-fold integration. Let us label the corresponding integration variables as $t_1,t_2,\ldots,t_{2n}$ from right to the left. Each shift $Z_i \to Z_i + t Z_j$ produced by ${\rm R}_{ij}$ induces the shift $\vec c_j \to \vec  c_j + t \vec c_i$ in the matrix $C_0$. Thus the sequence of R-operators in \p{Rn=RRPhi} and \p{Phin} brings us from $C_0$ to the top-cell of  ${\rm Gr}(2,n+2)$ parametrized with $t_1,\ldots,t_{2n}$. The measure $\omega(C)$ \p{omegaC2} takes the following form
\begin{align}
\omega(C)  = \frac{(t_{2n-2} t_{2n-3})^3}{\prod_{i=1}^{2n-4} t_i} \, \bigwedge_{i=1}^{2n} dt_i  \,. \label{}
\end{align}
which agrees with values of spectral parameters carried by R-operators \p{Rop2} in \p{Rn=RRPhi} and \p{Phin} .

\subsection{Cyclicity of the Yangian} 
\label{AppCycl}

The Yangian generators are entries of the monodromy matrix, which is a matrix product of the Lax operators. We will need the cyclic shift transformations of the Yangian generators, which we formulate in terms of the monodromy matrix.

If a homogeneous function $\ket{\Psi}$ of momentum twistors $Z_1,\ldots,Z_m$ is annihilated by the off-diagonal entries of the $4\times 4$ matrix formed as the matrix product of $m$ Lax operators \p{Lax2}
and the action of the diagonal entries results in an eigenvalue $\varphi(u,\delta)$,
\begin{align}
\left[ {\rm L}_1^{(\delta_1)}(u_1) \ldots {\rm L}_{m-1}^{(\delta_{m-1})}(u_{m-1}) \, {\rm L}_m^{(\delta_m)}(u_m) \right]^I{}_J \, \ket{\Psi} = \varphi(u,\delta)\, \delta^I_J \,\ket{\Psi}
\end{align}
then the matrix product of the cyclically shifted Lax operators has a similar action on $\ket{\Psi}$
\begin{align}
\left[ {\rm L}_m^{(\delta_m)}(u_m-4) \, {\rm L}_1^{(\delta_1)}(u_1) \ldots {\rm L}_{m-1}^{(\delta_{m-1})}(u_{m-1}) \right]^I{}_J \, \ket{\Psi} = \varphi'(u,\delta)  \, \delta^I_J \,\ket{\Psi}
\end{align}
with another eigenvalue $\varphi'(u,\delta)$,
\begin{align}
\varphi'(u,\delta) = \varphi(u,\delta) \cdot \frac{(u_m-4)(u_m -1+ \delta_m)}{u_m(u_m -1 + \delta_m) - \delta_m } \,.
\end{align}
One can check this statement following \cite{Chicherin:2013ora}.

\subsection{Lemmas}
\label{sec:lemmas}

Here we collect several technical statements which will be instrumental in  section ~\ref{AppRn}. 
In the following we deal with multiple products of the dual-conformal generators and we adopt the matrix notations,
\begin{align}
\left( \mathfrak{J}_a \, \mathfrak{J}_b \right) ^{I}{}_{J} := \left( \mathfrak{J}_a\right)^{I}{}_{K} \left( \mathfrak{J}_b\right)^{K}{}_{J} \,,\quad
\left( \mathfrak{J}_a \, \mathfrak{J}_b \, \mathfrak{J}_c \right) ^{I}{}_{J} := \left( \mathfrak{J}_a\right)^{I}{}_{K} \left( \mathfrak{J}_b\right)^{K}{}_{L} \left( \mathfrak{J}_c\right)^{L}{}_{J}  \,, \; \ldots
\end{align}

We observe that the products of several local dual-conformal generators $\mathfrak{J}_a$ \p{Jlocal} acting at the same spin-chain site labeled by $a$ are proportional to $\mathfrak{J}_a$,
\begin{align}
\left( \mathfrak{J}_a \, \mathfrak{J}_a \right)^{I}{}_{J} = (3+H_a) \, \left( \mathfrak{J}_a \right)^{I}{}_{J} \,, \label{JaJa}
\end{align}
and the proportionality coefficient can be considered as a number. Indeed, the dual-conformal weight counting operator $H_a$ \p{Hcount} commutes with the dual-conformal generators.
The latter equation enables us to invert the Lax operator \p{Lax2},
\begin{align}
\left[ \mathrm{L}^{(\delta)}_a (-u-3-\delta) \, \mathrm{L}^{(\delta)}_a (u) \right]^{I}{}_{J} = u (-u-3-\delta) \, \delta^I_J \,. \label{Linv}
\end{align}

Simplifications of the products of the dual-conformal generators $\mathfrak{J}_A$ and $\mathfrak{J}_B$ also appear when they act on functions of the twistor line $Z_A \wedge Z_B$, 
\begin{align}
Z_A \wedge Z_B : \qquad \left( \mathfrak{J}_A \mathfrak{J}_B \right)^{I}{}_{J} = - \left(\mathfrak{J}_A \right)^{I}{}_{J} \,,\quad  \qquad \left( \mathfrak{J}_B \mathfrak{J}_A \right)^{I}{}_{J} = - \left(\mathfrak{J}_B \right)^{I}{}_{J} \,. \label{JJtoLine}
\end{align}
In particular, we will need the product of the dual-conformal transformations $\bar{\mathfrak{J}}$ \p{JJbar}, which acts on the spin-chain sites $A$ and $B$, restricted to this space,
\begin{align}
Z_A \wedge Z_B : \qquad \left( \bar{\mathfrak{J}}\, \bar{\mathfrak{J}} \right)^{I}{}_{J} =  -2 \bar{\mathfrak{J}} ^{I}{}_{J}\,. \label{JJtoLine2}
\end{align}

A special role is played by the product of two R-operators which appear on the r.h.s. of \p{Rn=RRPhi}, namely ${\rm R}_{AB}^{(-1)} {\rm R}_{BA}^{(-1)}$. In view of \p{DCIRn2}, the product is invariant under the dual-conformal transformations with $\bar{\mathfrak{J}}$ \p{JJbar},
\begin{align}
\left[ {\rm R}_{AB}^{(-1)} {\rm R}_{BA}^{(-1)} , \bar{\mathfrak{J}}^{I}{}_{J} \right] = 0  \,, \label{RRJbar}
\end{align}
and it does not alter the dual-conformal weight
\begin{align}
\left[ H_a \, , \,  {\rm R}_{AB}^{(-1)} {\rm R}_{BA}^{(-1)}  \right] = 0 \;,\qquad a=A,B \,. \label{HRR}
\end{align}

Given an arbitrary function $G(Z_A,Z_B)$ with equal dual-conformal weights w.r.t. both twistor points $A$ and $B$,
\begin{align}
H_A\, G = H_B\, G \,, \label{HAG=HBG}
\end{align}
one can show that ${\rm R}_{AB}^{(-1)} {\rm R}_{BA}^{(-1)} G(Z_A,Z_B)$ is a function of the twistor line $Z_A \wedge Z_B$, namely it is invariant under shifts $Z_A \to Z_A+ \ep Z_B$ and $Z_B \to Z_B+ \ep Z_A$. Let us apply this observation to eq.~\p{Rn=RRPhi}. Indeed, $\ket{\Phi_n}$ depends on twistors $Z_A$ and $Z_B$, and it carries equal dual-conformal weights at points $A$ and $B$, see eq.~\p{RPhiweights}. Then $R_n$ given in \p{Rn=RRPhi} is a function of the line $Z_A \wedge Z_B$ in the twistor space.

The local exchange relations \p{RLL1} and \p{RLL2} result in the exchange relation satisfied by ${\rm R}_{AB}^{(-1)} {\rm R}_{BA}^{(-1)}$ with the product of Lax operators $\mathrm{L}_A\mathrm{L}_B$,
\begin{align}
 {\rm R}_{AB}^{(-1)} {\rm R}_{BA}^{(-1)} \, \left[ \mathrm{L}^{(-4)}_A(v+1) \mathrm{L}^{(-4)}_B(v) \right]^{I}{}_{J} = \left[ \mathrm{L}^{(-4)}_A(v) \mathrm{L}^{(-4)}_B(v+1) \right]^{I}{}_{J} \,  {\rm R}_{AB}^{(-1)} {\rm R}_{BA}^{(-1)} \,. \label{RARBLL}
\end{align}
Expanding the latter equation in powers of the spectral parameter $v$, we reproduce the statement of the dual-conformal invariance \p{RRJbar} as well as the following exchange relation, which is quadratic in the dual-conformal generators \p{Jlocal},
\begin{align}
 {\rm R}_{AB}^{(-1)} {\rm R}_{BA}^{(-1)} \left[ \mathfrak{J}_B +  \mathfrak{J}_A \mathfrak{J}_B \right]^{I}{}_{J} =   \left[ \mathfrak{J}_A +  \mathfrak{J}_A \mathfrak{J}_B \right]^{I}{}_{J}  {\rm R}_{AB}^{(-1)} {\rm R}_{BA}^{(-1)} \,. \label{RRmanyJ1}
\end{align}
In order to obtain more exchange relations which involve products of the dual-conformal generators, we contract the latter relation with the dual-conformal generators $\bar{\mathfrak{J}}$ \p{JJbar}, which commute with the product of the R-operators, see \p{RRJbar},
\begin{align}
 {\rm R}_{AB}^{(-1)} {\rm R}_{BA}^{(-1)} \left[ \bar{\mathfrak{J}} \, \mathfrak{J}_B +  \bar{\mathfrak{J}} \, \mathfrak{J}_A \mathfrak{J}_B \right]^{I}{}_{J} =   \left[ \bar{\mathfrak{J}} \, \mathfrak{J}_A +  \bar{\mathfrak{J}} \, \mathfrak{J}_A \mathfrak{J}_B \right]^{I}{}_{J}  {\rm R}_{AB}^{(-1)} {\rm R}_{BA}^{(-1)} \,. \label{RRmanyJ2}
\end{align}
In the following we work on the space of homogeneous functions carrying the dual-conformal weight equal to $-4$. The products of the dual-conformal generators from \p{RRmanyJ2} simplify as follows on this space  
\begin{align}
{\cal H}_A^{(-4)} \otimes {\cal H}_B^{(-4)} \; : \qquad  \begin{array}{l} \bar{\mathfrak{J}} \, \mathfrak{J}_B +  \bar{\mathfrak{J}} \, \mathfrak{J}_A \mathfrak{J}_B = -  \mathfrak{J}_B + \mathfrak{J}_B   \mathfrak{J}_A \mathfrak{J}_B \,,\\ 
\bar{\mathfrak{J}} \, \mathfrak{J}_A +  \bar{\mathfrak{J}} \, \mathfrak{J}_A \mathfrak{J}_B = -  \mathfrak{J}_A + \mathfrak{J}_B   \mathfrak{J}_A - \mathfrak{J}_A   \mathfrak{J}_B + \mathfrak{J}_B   \mathfrak{J}_A \mathfrak{J}_B \,, \end{array}
\end{align}
in view of \p{JaJa}. We take the sum of \p{RRmanyJ1} and \p{RRmanyJ2} and restrict the relation to the space of homogeneous functions, since the dual-conformal weights are not altered by $ {\rm R}_{AB}^{(-1)} {\rm R}_{BA}^{(-1)}$, see \p{HRR}, 
\begin{align}
{\cal H}_A^{(-4)} \otimes {\cal H}_B^{(-4)} \; : \qquad   {\rm R}_{AB}^{(-1)} {\rm R}_{BA}^{(-1)}  \left[ \mathfrak{J}_A \mathfrak{J}_B +  \mathfrak{J}_B \mathfrak{J}_A \mathfrak{J}_B \right]^{I}{}_{J} =  \left[ \mathfrak{J}_B \mathfrak{J}_A +  \mathfrak{J}_B \mathfrak{J}_A \mathfrak{J}_B \right]^{I}{}_{J} {\rm R}_{AB}^{(-1)} {\rm R}_{BA}^{(-1)} \label{RRmanyJ3}
\end{align}
According to the discussion around eq.~\p{HAG=HBG}, the r.h.s. of \p{RRmanyJ3} is a function of the twistor line $Z_A \wedge Z_B$. The product of the dual-conformal generators $\mathfrak{J}_B \mathfrak{J}_A +  \mathfrak{J}_B \mathfrak{J}_A \mathfrak{J}_B$ from the r.h.s. of \p{RRmanyJ3} vanishes on this space, see \p{JJtoLine}. Finally, after rewriting the product of the dual-conformal generators in the l.h.s. of \p{HAG=HBG}, 
\begin{align}
{\cal H}_A^{(-4)} \otimes {\cal H}_B^{(-4)} \; : \qquad 
\mathfrak{J}_A \mathfrak{J}_B +  \mathfrak{J}_B \mathfrak{J}_A \mathfrak{J}_B = 2 \mathfrak{J}_A \mathfrak{J}_B  + \bar{\mathfrak{J}}\, \mathfrak{J}_A \mathfrak{J}_B,
\end{align}
eq.~\p{HAG=HBG} takes the following form  
\begin{align}
{\cal H}_A^{(-4)} \otimes {\cal H}_B^{(-4)} \; : \qquad   {\rm R}_{AB}^{(-1)} {\rm R}_{BA}^{(-1)}  \left[   2 \mathfrak{J}_A \mathfrak{J}_B  + \bar{\mathfrak{J}}\, \mathfrak{J}_A \mathfrak{J}_B \right]^{I}{}_{J} = 0 \,. \label{RRmanyJ=0}
\end{align}

The relations established in this Section are required in the next Section.

\subsection{Yangian invariances of $R_n$}
\label{AppRn}

We are going to establish how the Yangian generators from the monodromy matrix expansion \p{Texpand2} act on the leading singularities $R_n$. To this end, we form an auxiliary monodromy matrix out of the Lax operators \p{Lax2}
taken in the order $A,1,2,\ldots,n,B$. The auxiliary monodromy matrix acts diagonally on the pseudo-vacuum state $
\ket{\Omega_n}$ \p{vac2},
\begin{align}
\left[ {\rm L}_A^{(0)} (u) {\rm L}_{1}^{(-4)}(u) {\rm L}_{2}^{(0)}(u) \ldots {\rm L}_{n-1}^{(0)}(u) {\rm L}_n^{(-4)}(u+4) {\rm L}_{B}^{(0)}(u) \right]^{I}{}_{J} \ket{\Omega_n} = u^n (u-1)(u+3)\, \delta^I_J \, \ket{\Omega_n} \,, \label{TauxOm}
\end{align}
since each of the constituent Lax operators acts diagonally on the delta-functions, see \p{Lax2}, 
\begin{align}
& \left[ {\rm L}_i (u) \right]^{I}{}_{J} \, \delta^{4}(Z_j) = u \,\delta^I_J\,  \delta^4(Z_j) \;, \quad i \neq j \\
& \left[ {\rm L}_i (u) \right]^{I}{}_{J} \, \delta^{4}(Z_i) = (u-1) \,\delta^I_J\,  \delta^4(Z_i) \,.
\end{align}
Then we act with $(2n-2)$ R-operators from \p{Phin}
onto \p{TauxOm} and pull them through the products of Lax operators by means of the local exchange relations \p{RLL1} and \p{RLL2},
\begin{align}
\left[ {\rm L}^{(-4)}_A(u) {\rm L}_{1}^{(0)}(u) {\rm L}_{2}^{(0)}(u) \ldots {\rm L}_{n-1}^{(0)}(u) {\rm L}_n^{(0)}(u) {\rm L}_{B}^{(-4)}(u+4) \right]^{I}{}_{J} \ket{\Phi_n} = u^n (u-1)(u+3)\, \delta^I_J \, \ket{\Phi_n} \,.
\end{align}
We apply the cyclicity of  App.~\ref{AppCycl} to the previous monodromy relation cyclically shifting the ordering of the Lax operators from $A,1,2,\ldots,n,B$ to $B,A,1,2,\ldots,n$ and replacing the monodromy matrix $\mathrm{L}_1 \ldots \mathrm{L}_n$ with the shorthand notation \p{Tmonod2},
\begin{align}
\left[ {\rm L}_{B}^{(-4)}(u) {\rm L}_A^{(-4)}(u) {\rm T}(u)  \right]^{I}{}_{J} \ket{\Phi_n} = u^n (u-1)^2 \, \delta^I_J \, \ket{\Phi_n} \,, \label{monodPhi}
\end{align}
and we invert $\mathrm{L}_A$ and $\mathrm{L}_B$ on the l.h.s. of the previous equation by means of \p{Linv},
\begin{align}
\left[ {\rm T}(u) \right]^I{}_J \, \ket{\Phi_n}  = u^{n-2} \left[ \mathrm{L}^{(-4)}_A(1-u)\mathrm{L}_B^{(-4)}(1-u)\right]^{I}{}_{J} \ket{\Phi_n} \,. \label{TPhin}
\end{align}

We need to act with ${\rm R}_{AB}^{(-1)} {\rm R}_{BA}^{(-1)}$ on both sides of \p{TPhin} in order to promote $\ket{\Phi_n}$ to the leading singularities $R_n$ according to \p{Rn=RRPhi}. This is straightforward in the l.h.s. of \p{TPhin}. If we could pull this R-operator product through $\mathrm{L}_A \mathrm{L}_B$ in the r.h.s. of \p{TPhin} then we would find that $R_n$ is invariant under all Yangian generators. However, the latter is not true. The obstacle is that values of spectral parameters in $\mathrm{L}_A \mathrm{L}_B$ in the r.h.s. of \p{TPhin} are not adjusted like in \p{RARBLL}.

In order to obtain symmetry relations for $R_n$, we expand \p{TPhin} in the spectral parameter taking into account definitions in \p{Texpand2}, \p{JJbar} and \p{Jhat2},
\begin{align}
& u^{n-1}\;  : \quad \mathfrak{J}^{I}{}_{J} \ket{\Phi_n} = - \bar{\mathfrak{J}}^{I}{}_{J}\ket{\Phi_n} - 2 \delta^{I}_{J} \ket{\Phi_n} \,, \label{un-1} \\
& u^{n-2}\;  : \quad \widehat{\mathfrak{J}}^{I}{}_{J} \ket{\Phi_n} =  \delta^{I}_{J} \ket{\Phi_n} + \bar{\mathfrak{J}}^{I}{}_{J}\ket{\Phi_n} + \left( \mathfrak{J}_A \mathfrak{J}_B  \right)^{I}{}_{J} \ket{\Phi_n}\,, \label{un-2} \\
& u^{n-1-k}\;  : \quad  \left( \mathfrak{J}^{(k)} \right)^{I}{}_{J} \ket{\Phi_n} = 0 \;,\quad k =2,\ldots,n-1 \,. \label{un-1-k}
\end{align}
We act with ${\rm R}_{AB}^{(-1)} {\rm R}_{BA}^{(-1)}$ onto \p{un-1} and \p{un-1-k}
and replace there $\ket{\Phi_n}$ with $R_n$ according to \p{Rn=RRPhi}. Indeed, the R-operators are dual-conformal invariant, see eq.~\p{RRJbar}. Thus, \p{un-1} is equivalent to the dual-conformal invariance of the leading singularities $R_n$ \p{DCIRn2}, while \p{un-1-k} is equivalent to the Yangian invariance of $R_n$ with the level-two and higher-level Yangian generators \p{JkRn}.  
Only \p{un-2} cannot be immediately interpreted as the Yangian invariance relation, since ${\rm R}_{AB}^{(-1)} {\rm R}_{BA}^{(-1)}$ does not commute nicely with $\mathfrak{J}_A \mathfrak{J}_B$ in the last term on the r.h.s. of \p{un-2}.

Upon contraction of \p{un-2} with the tensor $\vev{Z_A,Z_{B},*,\star}$ the unwanted terms drop out,
\begin{align}
\ep_{I K L M} Z^K_A Z^L_{B} \,  \widehat{\mathfrak{J}}^{I}{}_{J} \, \ket{\Phi_n} =  \ep_{J K L M} Z^K_A Z^L_{B} \, \ket{\Phi_n} \,,
\end{align}
and the latter relation is equivalent to \p{YangianRn} since both ${\rm R}_{AB}$ and ${\rm R}_{BA}$, which act as shifts on twistors $Z_A$ and $Z_B$, commute with $\vev{Z_A,Z_{B},*,\star}$.

Another way to acquire a symmetry relation with the level-one Yangian generators $\widehat{\mathfrak{J}}$ from \p{un-2} is to cook up a polynomial in the local dual-conformal generators $\mathfrak{J}_A$ and $\mathfrak{J}_B$ which nicely commutes with ${\rm R}_{AB}^{(-1)} {\rm R}_{BA}^{(-1)}$. We have established in \p{RRmanyJ=0} that this is the case for $\bar{\mathfrak{J}} \, \mathfrak{J}_A \mathfrak{J}_B + 2\mathfrak{J}_A \mathfrak{J}_B$. Indeed, contracting \p{un-2} with $\bar{\mathfrak{J}}^{K}{}_{I} + 2\delta^{K}_{I}$ we organize the unwanted terms into this polynomial,
\begin{align}
\left( \bar{\mathfrak{J}}\, \widehat{\mathfrak{J}} + 2 \widehat{\mathfrak{J}} \right)^{I}{}_{J}  \ket{\Phi_n} = 2 \delta^{I}_{J} \ket{\Phi_n} + 3 \bar{\mathfrak{J}}^{I}{}_{J}\ket{\Phi_n} + \left(\bar{\mathfrak{J}}\, \bar{\mathfrak{J}} \right)^{I}{}_{J}\ket{\Phi_n}  + \left( \bar{\mathfrak{J}} \, \mathfrak{J}_A \mathfrak{J}_B + 2\mathfrak{J}_A \mathfrak{J}_B \right)^{I}{}_{J} \ket{\Phi_n} \,.
\end{align}
We act on the previous relation with ${\rm R}_{AB}^{(-1)} {\rm R}_{BA}^{(-1)}$ pulling it through $\bar{\mathfrak{J}}$ \p{RRJbar}, replacing $\bar{\mathfrak{J}}\, \bar{\mathfrak{J}} \to -2 \bar{\mathfrak{J}}$
according to \p{JJtoLine2}, and throwing out the last term on the r.h.s. in view of \p{RRmanyJ=0},
\begin{align}
\left( \bar{\mathfrak{J}}\, \widehat{\mathfrak{J}} + 2 \widehat{\mathfrak{J}} \right)^{I}{}_{J}  \, R_n = 2 \delta^{I}_{J} \, R_n + \bar{\mathfrak{J}}^{I}{}_{J} \, R_n\,.
\end{align}
The latter is the symmetry relation \p{JbarJhat}.

\bibliographystyle{JHEP}
\bibliography{wl5plus1.bib,publications.bib}

\end{document}